\pdfoutput=1
\documentclass[cernpreprint=true, atlasdraft=false, texlive=2023, UKenglish, coverpage=false, texmf, orcidlogo]{atlasdoc}
 
\usepackage[subfigure]{atlaspackage}
\usepackage{atlaspackage}
\usepackage{pdflscape}
 
\usepackage{atlasbiblatex}
 
\usepackage{makecell}
 
\usepackage{atlasphysics}
\usepackage{atlasjetetmiss}
\usepackage{atlasjournal}

\graphicspath{{logos/}{figures/}}
 
\usepackage{ANA-STDM-2020-20-PAPER-defs}

\AtlasTitle{Measurements of multijet event isotropies using optimal transport with the ATLAS detector}

\AtlasAbstract{
A measurement of novel event shapes quantifying the isotropy of collider events is performed in 140~fb$^{-1}$~of proton--proton collisions with $\sqrt{s}=13$~TeV centre-of-mass energy recorded with the ATLAS detector at CERN's Large Hadron Collider.
These event shapes are defined as the Wasserstein distance between collider events and isotropic reference geometries.
This distance is evaluated by solving optimal transport problems, using the `Energy-Mover's Distance'.
Isotropic references with cylindrical and circular symmetries are studied, to probe the symmetries of interest at hadron colliders.
The novel event-shape observables defined in this way are infrared- and collinear-safe, have improved dynamic range and have greater sensitivity to isotropic radiation patterns than other event shapes.
The measured event-shape variables are corrected for detector effects, and presented in inclusive bins of jet multiplicity and the scalar sum of the two leading jets' transverse momenta.
The measured distributions are provided as inputs to future Monte Carlo tuning campaigns and other studies probing fundamental properties of QCD and the production of hadronic final states up to the TeV-scale.
}

\AtlasRefCode{STDM-2020-20}
 
\PreprintIdNumber{CERN-EP-2023-079}

\arXivId{2305.16930}
 
\HepDataRecord{2663035}
 
\AtlasJournalRef{\JHEP 10 (2023) 060}
\AtlasDOI{10.1007/JHEP10(2023)060}

\hypersetup{pdftitle={ATLAS document},pdfauthor={The ATLAS Collaboration}}
 
\begin{document}
 
\maketitle
 
\tableofcontents

\section{Introduction}

Event shapes are a family of observables used to describe the flow of energy in collider events~\cite{Banfi:2010xy}.
Measurements of event shapes have been used to probe fundamental properties of QCD~\cite{PhysRevLett.39.1587, TASSO:1979zyf, BARTEL1980142, TASSO:1983cre, BARBER1979139, MARK-II:1989aml, MARK-II:1989aml, L3:1990jlf, L3:1992btq,TASSO:1990cdg, SLD:1994idb, ALEPH:1995njx, Kluth:2000km, OPAL:2001klt, ALEPH:2003obs, DELPHI:2003yqh, Dasgupta:2003iq, L3:2004cdh, OPAL:2004wof,  H1:2005zsk, ZEUS:2006vwm, Kluth:2009nx, OPAL:2011aa, Abbiendi:2008zz, Tanabashi:2018oca, Nason:2023asn}, to tune Monte Carlo (MC) models~\cite{Skands:2014pea,ATL-PHYS-PUB-2014-021,CMS-GEN-17-001} and to search for physics beyond the Standard Model (SM)~\cite{Anchordoqui:2010hi,Krauss:2016ely,Goldouzian:2020wdq,Buckley:2020wzk,Buckley:2021neu}.
A novel class of event shape observables was recently proposed to quantify the isotropy of collider events~\cite{Cesarotti:2020hwb}.
These observables, broadly called \emph{event isotropy}, measure how `far' a collider event is from a symmetric radiation pattern in terms of a Wasserstein distance metric~\cite{wasserstein1969markov, dobrushin1970prescribing}.
Event isotropy observables are complementary to canonical event shapes such as thrust~\cite{PhysRevLett.39.1587,BRANDT196457,DERUJULA1978387}, sphericity~\cite{PhysRevD.1.1416,Ellis:1976uc} and spherocity~\cite{PhysRevLett.39.1237}, which were designed to quantify how closely collider events resemble `pencil-like' dijet events.
 
The Wasserstein distances used to define event isotropies are framed in terms of optimal transport problems, using the `Energy-Mover's Distance' (EMD)~\cite{Komiske:2019fks, Komiske:2020qhg}.
This formulation is infrared- and collinear-safe by construction, avoiding pathologies related to low-energy particles or small-angled splittings that can affect other event shapes.
Event isotropies are more sensitive to isotropic radiation patterns than other event shapes, isolating events with larger multiplicities of objects that are isotropically distributed.
This behaviour differs from that of the thrust and sphericity, which interpolate between back-to-back dijet and well-balanced trijet events.
Event isotropies also exhibit larger dynamic ranges in the quasi-isotropic event-shape region than traditional event shapes~\cite{Cesarotti:2020ngq}.
This distinct behaviour implies that event isotropy observables have the potential to be powerful references for Monte Carlo developers, complementary to existing measurements.
They also have the potential to increase the sensitivity of many measurements and searches for rare processes within and beyond the Standard Model with isotropically distributed signals and backgrounds, such as $\ttbar H$~\cite{HIGG-2020-03,CMS-HIG-21-006} or $\ttbar\ttbar$~\cite{TOPQ-2021-08,CMS-TOP-22-013} production or searches for QCD instantons~\cite{Khoze:2019jta,Amoroso:2020zrz}, strongly-coupled hidden sectors~\cite{Knapen:2016hky,Cesarotti:2020uod}, models with large extra dimensions~\cite{Arkani-Hamed:1998jmv,Antoniadis:1998ig,Arkani-Hamed:1998sfv,Randall:1999ee,Randall:1999vf} or microscopic black holes~\cite{Giddings:2001bu,Anchordoqui:2002fc,Landsberg:2006mm,Meade:2007sz,Dimopoulos:2001hw}.
 
ATLAS and CMS have performed differential measurements of many canonical event-shape observables at the LHC in minimum-bias~\cite{CMS-QCD-10-013,STDM-2011-32}, multijet~\cite{STDM-2011-33,CMS-SMP-17-003,STDM-2019-02} and $Z$+jet~\cite{CMS-EWK-11-021,STDM-2014-07} final states. Earlier hadron-collider event-shape measurements were also performed by the CDF~\cite{CDF:1991etg} and D{\O}~\cite{Sorin:2003xj} collaborations.
The large \RunTwo dataset~\cite{DAPR-2018-01} and advances in jet reconstruction performance~\cite{JETM-2018-05} and MC modelling~\cite{ATL-PHYS-PUB-2019-017} allow the latest event-shape measurements to be made with fine binning and  differentially in inclusive bins of jet multiplicity, $\njets$, and the scalar sum of the leading and subleading jet transverse momenta, $\HTTwo=p_{\text{T},1}+p_{\text{T},2}$.
 
This paper presents normalised differential cross-section measurements of three event-isotropy observables, introduced in detail in Section~\ref{sec:intro:iso}.
These measurements allow the shape of the event isotropy observables to be studied in detail; particularly by making comparisons with predictions from several cutting-edge Monte Carlo event generators.
The methodology closely follows that used in the most recent ATLAS measurement of canonical event shapes~\cite{STDM-2019-02}, using \akt jets with radius parameter $R=0.4$ as the input objects for event shape calculations (Section~\ref{sec:method:jets})~\cite{Cacciari:2008gp}.
 
The structure of this paper is as follows. After these introductory remarks, an overview of the central phenomenological concepts used in this analysis is provided.
Section~\ref{sec:intro:iso} defines the three event-isotropy observables that are later measured, describing how they are calculated and some of their general properties.
A description of the ATLAS detector, the \RunTwo dataset and the simulated multijet events used in this analysis may be found in Section~\ref{sec:atlas}.
In Section~\ref{sec:methodology}, details of the physics object reconstruction, event selection and unfolding procedure used in this analysis are given.
A summary of the systematic uncertainties considered in the measurement is provided in Section~\ref{sec:systematics}.
The main results of the analysis are presented in Section~\ref{sec:results}; afterwards, concluding remarks are made.
 
\FloatBarrier
\subsection{The Energy-Mover's Distance}\label{sec:intro:emd}
 
In order to compute how `far' one event is from another, a well-defined mathematical definition of distance must be introduced.
Event isotropy is computed using the Energy-Mover's Distance~(EMD)~\cite{Komiske:2019fks,Komiske:2020qhg} -- an application of the well-known `Earth-Mover's Distance' from computer vision~\cite{10.1007/978-3-642-40020-9_43,192468, Rubner:1998:MDA:938978.939133,Rubner2000,10.1007/978-3-540-88690-7_37} to particle physics, using the $p$-Wasserstein metric~\cite{wasserstein1969markov, dobrushin1970prescribing}.
The EMD is defined as \emph{the minimum amount of `work' necessary to transport one event $\mathcal{E}$ with $M$ particles into another $\mathcal{E^{\prime}}$ of equal energy with $M'$ particles, by movements of energy $f_{ij}$ from particle $i\leq M$ in one event to particle $j\leq M'$ in the other:}\footnote{In this analysis, the energies of the events compared in each EMD calculation are always normalised to each other, and so the EMD is presented here in a simplified form.}
 
\begin{equation}
\label{eq:emd}
\EMD_{\beta} (\mathcal E, \mathcal E') = \min_{\{f_{ij}\ge0\}} \sum_{i=1}^M\sum_{j=1}^{M'} f_{ij} \theta_{ij}^\beta,
\end{equation}
\begin{equation}
\label{eq:emdconstraints}
\sum_{i=1}^M f_{ij} = E_j', \quad\quad \sum_{j=1}^{M'} f_{ij} = E_i, \quad\quad \sum_{i=1}^M\sum_{j=1}^{M'} f_{ij} = \sum_{i=1}^{M} E_i = \sum_{j=1}^{M'} E_j = E_\text{total}
\end{equation}
where $\theta_{ij}$ is a pairwise distance between particles known as the \emph{ground measure}, $\beta > 0$ is an angular weighting exponent, and $E_\text{total}$ is the total energy in each event.
The constraints defined in Eq.~(\ref{eq:emdconstraints}) ensure that the amount of energy moved from a particle is positive and does not exceed its initial energy, and that the total energy moved is conserved before and after the transport operation.
 
Equations~(\ref{eq:emd}) and~(\ref{eq:emdconstraints}) define an \emph{optimal transport} problem between the energy flow in events $\mathcal E$ and $\mathcal E'$, which may be solved using common scientific computing libraries.
The open-source packages \AtlasMC{event\_isotropy}~\cite{eventiso} (which implements event isotropy calculations using the \AtlasMC{Python Optimal Transport} (\AtlasMC{POT}) library~\cite{flamary2021pot}) and \AtlasMC{Wasserstein}~\cite{Komiske:2019fks,Komiske:2020qhg,10.1145/2070781.2024192} were both tested during this analysis, and were found to give compatible results.
 
In these measurements, the input objects to EMD calculations are \akt jets with radius parameter $R=0.4$ (reconstructed and calibrated as described in Section~\ref{sec:method:jets})~\cite{Cacciari:2008gp}.
The use of jets rather than per-particle inputs results in efficient optimal transport calculations because of the lower resultant object multiplicity per-event, and reduces sensitivity to non-perturbative QCD effects such as hadronisation.
Experimentally, this choice provides a well-calibrated object with a precisely measured energy scale and energy resolution (Section~\ref{sec:systematics:jesjer}) that has a clearly defined counterpart at particle level for use in the unfolding procedure (Section~\ref{sec:method:unfolding}), and a measurement that is infrared- and collinear-safe.
 
\FloatBarrier
\subsection{Event shapes via optimal transport}
 
References~\cite{Cesarotti:2020hwb} and~\cite{Komiske:2020qhg} interpret well-established event shapes such as the thrust and spherocity in terms of the geometric approach made possible by establishing a distance metric between different radiation configurations.
This discussion is summarised in this section, but interested readers are referred to the original publications for a more thorough discussion.
 
\begin{figure}[htbp]
\centering
\includegraphics[width=0.66\textwidth]{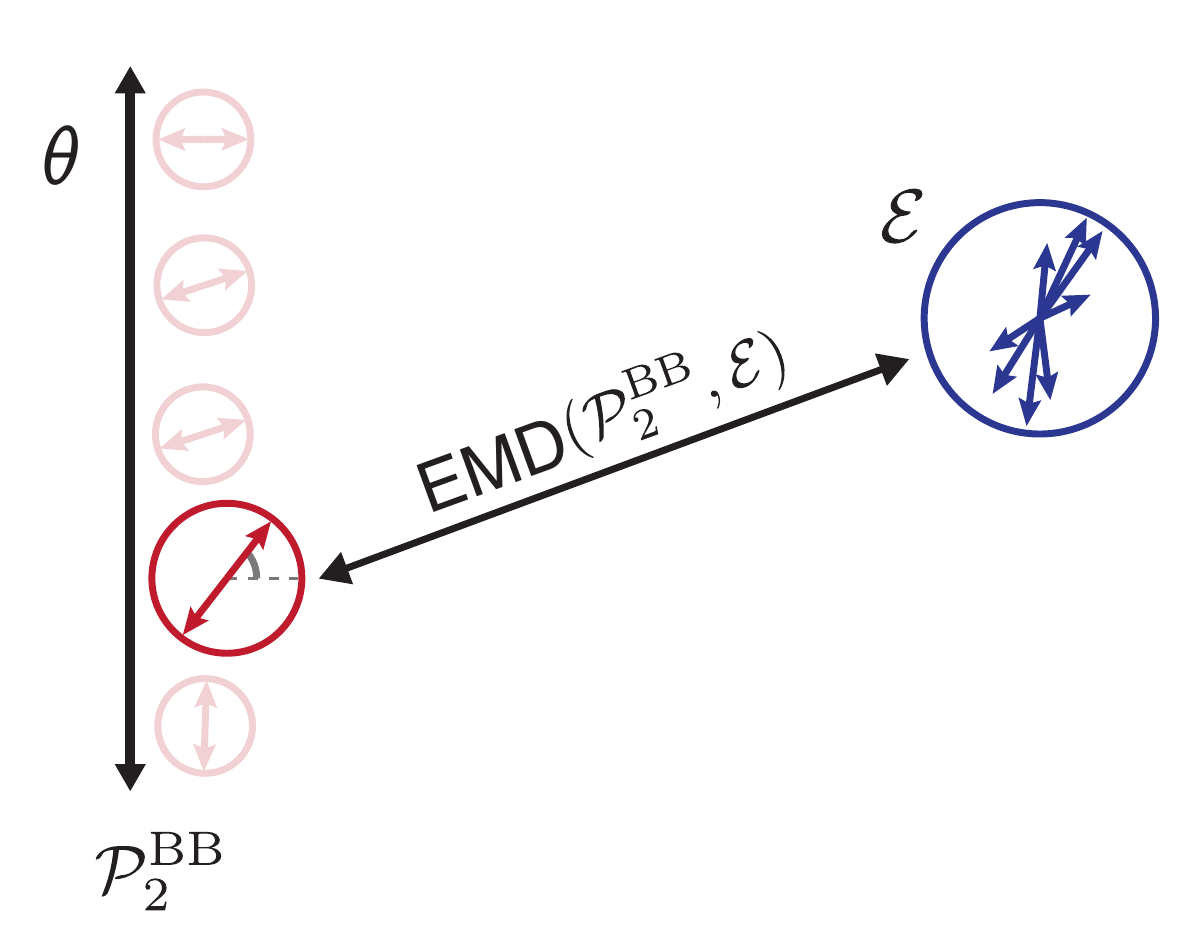}
\caption{A representation of the EMD as the minimum distance of an event $\mathcal{E}$ to a manifold of two-particle back-to-back events $\mathcal{P}_{2}^{BB}$.
The angle $\theta$ represents the degree of freedom corresponding to the relative azimuthal orientation of the reference event from $\mathcal{P}_{2}^{BB}$ and the collider event $\mathcal{E}$.
The arrow indicates, schematically, the EMD describing the minimum cost to transport $\mathcal{E}$ to $\mathcal{P}_{2}^{BB}$.
}
\label{fig:geometry:c}
\end{figure}

Some canonical event-shape observables can be identified as the minimum EMD between a collider event and a manifold defined by a certain radiation pattern.
For example, the simplest case considers the $\mathcal P^{\text{BB}}_2$ manifold, defined by the set of all back-to-back two-particle events with energy $E^i_\text{BB}$:
\begin{equation*}
\mathcal P^{\text{BB}}_2 = \left\{\left. \sum_{i=1}^2 E^i_\text{BB}\, \delta(\hat n - \hat n_i)\,\, \right| \,\, E^i_\text{BB} \ge 0, \,\, \hat{n}_1 = - \hat{n}_2 \right\}.
\end{equation*}
where $\hat n$ is along the direction of the thrust axis, and $\hat n_i = \vec p_i/E_\text{BB}$, and $\vec p_i$ is the momentum of particle $i$.
The event thrust may be constructed in terms of an optimal transport problem between the collider event $\mathcal{E}$ and $\mathcal{P}^{\text{BB}}_{2}$.\footnote{When computing the transverse thrust, the two energies do not have to be equal.}
A common definition of the thrust for an event with $M$ massless particles and total energy $E_\text{total}$ is given by
\begin{align}
t(\mathcal{E}) &= 2\min_{\hat n}\sum_{i=1}^M\frac{|\vec p_i|(1- |\vec n_i \cdot \hat n|)}{E_\text{total}} \nonumber \\
&= 2\min_{\hat n}\sum_{i=1}^M \frac{E_i}{E_\text{total}} \min(1 - \hat n_i \cdot \hat n,\, 1 + \hat n_i \cdot \hat n).\label{eqn:thrust}
\end{align}
From Eq.~(\ref{eqn:thrust}), it is clear that thrust can be formulated as the transportation cost to move particle $i$ to either $\hat{n}$ or $-\hat{n}$, with an angular measure of $$\theta_{ij}^2 = 2n_i^\mu n_{j\mu}= 2 (1 - |\vec n_i \cdot \hat n|)$$ and a normalised energy weight $$f_{ij} = \frac{|\vec p_i|}{E_\text{total}}.$$
The minimisation over $\hat{n}$ is equivalent to finding the thrust axis of the event.
This expression is identified as the EMD between $\mathcal{E}$ and the closest event $\mathcal{E}^{\prime}\in\mathcal{P}^{\text{BB}}_{2}$:
\begin{equation*}
t(\mathcal E) = \text{EMD}(\mathcal{E}, \mathcal{E}^{\prime} ),
\end{equation*}
with $\beta=2$ in the notation of Eq.~(\ref{eq:emd}).
The transverse thrust is obtained by making a transverse projection of the events and considering only the ring of back-to-back particle geometries (or, `dipole-like geometries') made by the subset of $\mathcal{P}^{\text{BB}}_{2}$ which exists in the transverse plane (illustrated in Figure~\ref{fig:geometry:c}).
 
Similarly, choosing $\beta=1$ instead yields the event spherocity, while considering the larger manifold of all two-particle configurations $\mathcal{P}_{2}$ yields the event broadening~\cite{Catani:1992jc}.
The smallest distances between the sets of $N$-particle manifolds (which do not necessarily conserve momentum) and $\mathcal{E}$ are equivalent to the event $N$-jettinesses~\cite{Stewart:2010tn}.\footnote{Analogously, performing such a calculation within a jet instead of an event results in the $N$-subjettiness~\cite{Thaler:2010tr}.}
 
Following the example of these event isotropy variables, other event shapes could be constructed to probe different aspects of QCD radiation in collider events.
The development of other observables with potentially model- or search-specific applications, e.g.\ those with non-jetty geometries~\cite{Strassler:2008bv,Knapen:2016hky}, may be a promising area for further study.
 
\FloatBarrier
\section{Event isotropies}\label{sec:intro:iso}
 
The `event isotropy' $\Iso$~\cite{Cesarotti:2020hwb} builds upon the set of event shapes defined as distances from finite-particle configurations.
This observable is defined as a Wasserstein distance between a collider event $\mathcal{E}$ and a (quasi-)uniform radiation pattern $\mathcal{U}$, determined using the Energy-Mover's Distance (Section~\ref{sec:intro:emd}):
\begin{equation*}
\mathcal{I} \left(\mathcal{E}\right) = \text{EMD}\left(\mathcal{E}, \mathcal{U}\right),
\end{equation*}
The total energy in each reference event $\mathcal{U}$ is defined to be the same as that for the collider event $\mathcal{E}$ it is compared with, so that the EMD is computed with normalised energy transfer ($f_{ij} \rightarrow f_{ij}/E_\text{total}$).
This ensures that the event isotropy \Iso\ is bounded on $\Iso\in[0,1]$ and is dimensionless.
The least isotropic events take values approaching $\mathcal{I}=1$.
By construction, perfectly (and only perfectly) isotropic events take a value of $\mathcal{I}=0$, meaning there is zero distance between radiation patterns.
Event isotropies can therefore differentiate between quasi-isotropic events better than the existing observables $C$-parameter and $D$-parameter~\cite{ELLIS1981421}, which respectively take extreme values for events with symmetric radiation along three perpendicular axes and events that are planar.
 
Event isotropy observables are defined on sets of massless input particles with no net transverse momentum.
In this study, the input particles are the event's reconstructed \akt~jets at either detector level or particle level (Section~\ref{sec:method:jets}).
While these objects are massive, only their transverse momentum (\pt) and angular kinematic information (rapidity, $y$, and azimuthal angle, $\phi$) are used for the isotropy calculation.
The recoil term is added to $\mathcal{E}$ before computing the isotropy, following the description in Ref.~\cite{Cesarotti:2020hwb}, to ensure the resulting distribution of \Iso\ is bounded.
This quantity is computed as the negative four-vector sum of all jets inputted to the EMD calculation for a given event.
 
Different choices of the reference event $\mathcal{U}$ can be made, possessing alternative geometrical symmetries.
Observables with both one-dimensional `ring-like' and two-dimensional `cylindrical' symmetries are studied in this analysis.
In practice, any reference geometry must be constructed using a user-defined finite number of particles, $N$, for the EMD to be computed using numerical methods.
This parameter should be chosen such that it is large enough to maintain approximately continuous symmetries while balancing this against the computational expense, since the complexity of optimal transport problems with $n$ particles scales naively as $\mathcal{O}(n^3 \log^2 n)$.
The minimal choice of $N$ that prevented discretisation effects was used in this analysis to facilitate this analysis of the large \RunTwo LHC dataset.
 
\begin{figure}[htb]
\centering
\subfigure[]{\includegraphics[width=0.32\textwidth]{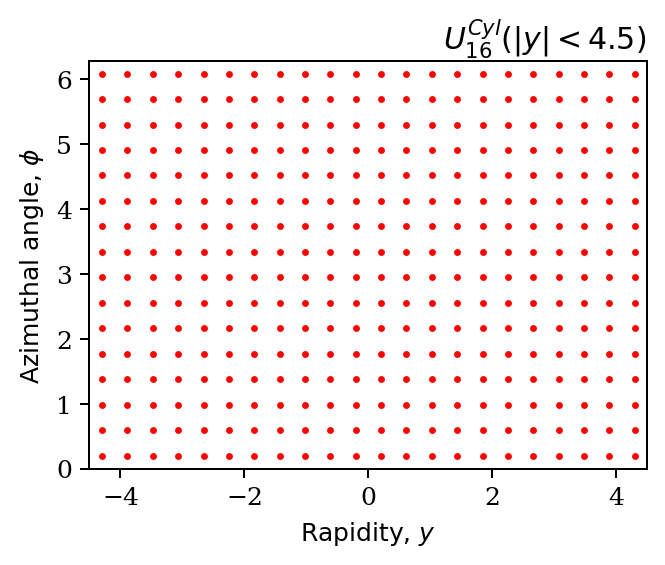}}
\subfigure[]{\includegraphics[width=0.32\textwidth]{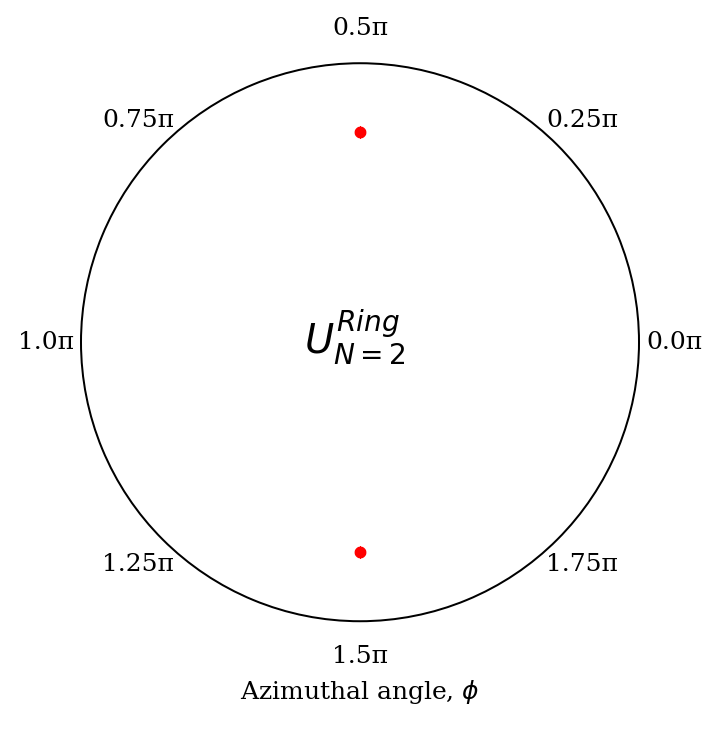}}
\subfigure[]{\includegraphics[width=0.32\textwidth]{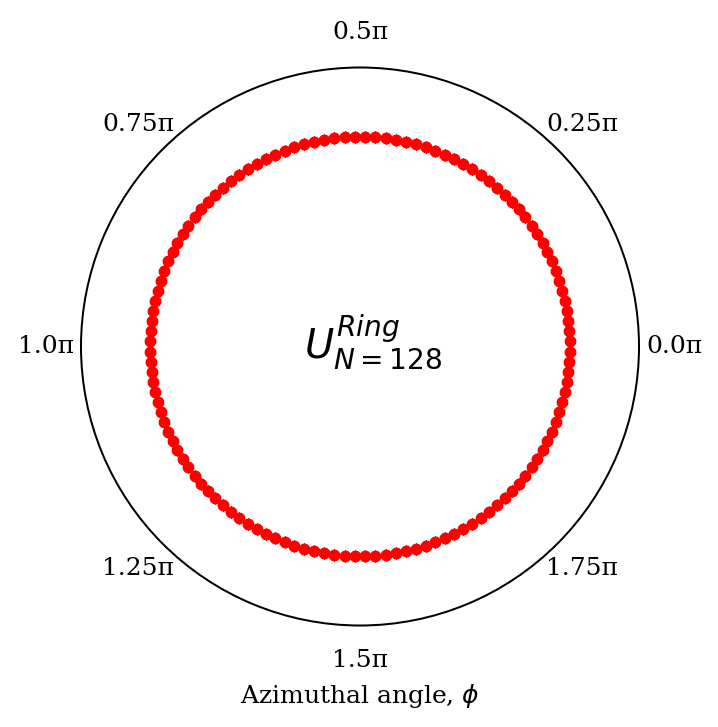}}
\caption{
Reference geometries with (a) cylindrical and ring-like symmetry with (b) 2 and (c) 128 reference points.
The radius at which the points in the ring-like geometry are located is arbitrary.
}
\label{fig:intro:refgeos}
\end{figure}

\begin{figure}[htb]
\centering
\includegraphics[width=0.95\textwidth]{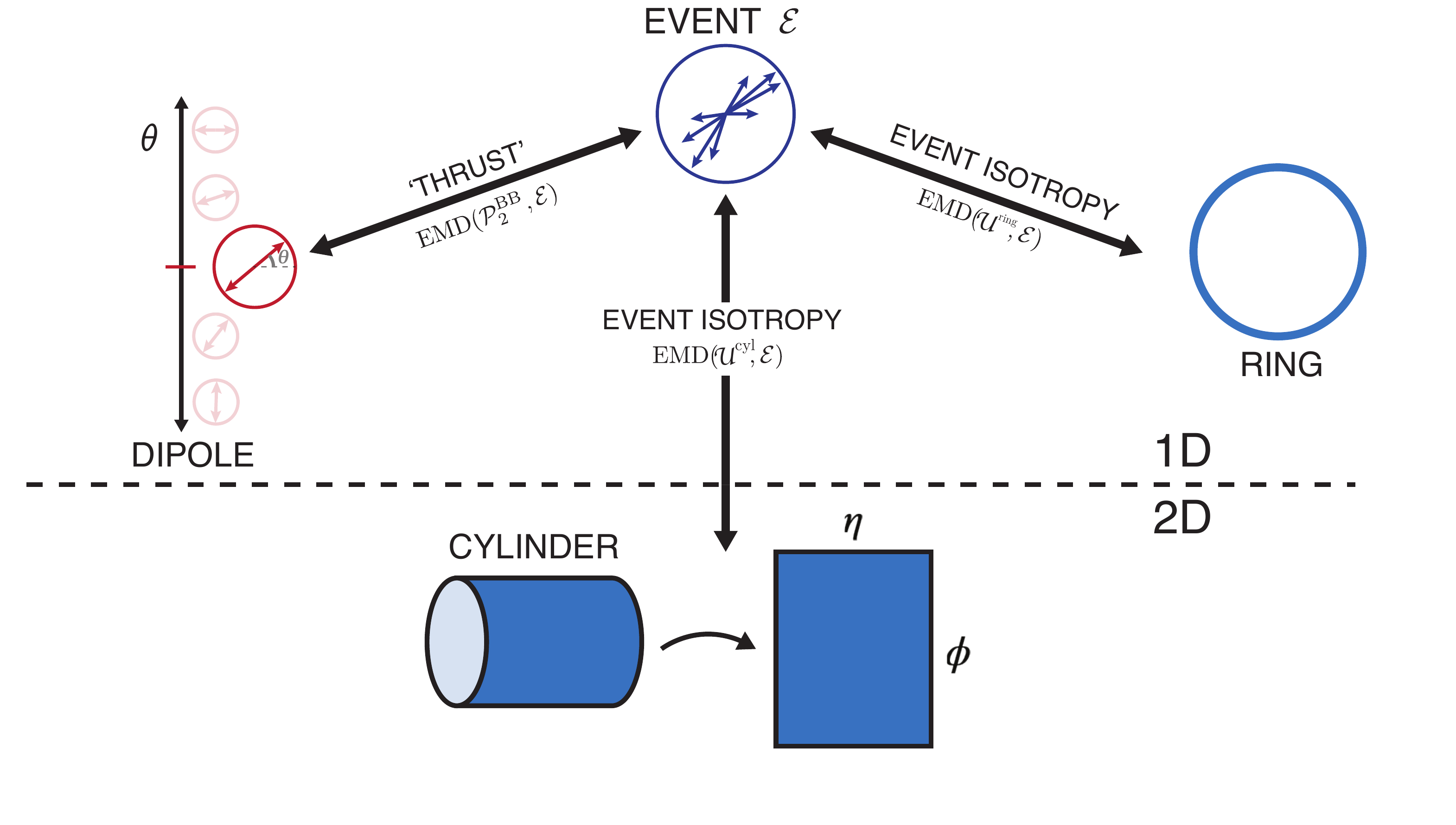}
\caption{
Schematic illustrating the three observables measured in this analysis in terms of the Energy-Mover's Distance from a collider event, $\mathcal{E}$: ~$\IsoRingTwo$,~$\IsoCylSixteen$,~and~$1-\IsoRingOneTwentyEight$.
}\label{fig:geometry:carifigure2}
\end{figure}

Three event-shape observables are considered in this analysis.
Both a quasi-uniform ring-like geometry with $N=128$ points (\IsoRingOneTwentyEight), and the special case of a ring-like geometry with $N=2$ (\IsoRingTwo) are studied.
This $N=2$ observable is similar to transverse thrust, but with balanced energy as mentioned earlier.
These two cases are studied to more directly compare the behaviours of the isotropy observables because they are defined on sets with zero net transverse momentum, unlike thrust.
A quasi-uniform cylindrical geometry with $N=16$ azimuthal segments is also considered (\IsoCylSixteen), resulting in a square reference grid of 352 points in the event rapidity--azimuth plane of $y\in[-4.5,4.5]$ and $\phi\in[0,2\pi]$ (matching the acceptance region of $R=0.4$ jets in ATLAS, described in Section~\ref{sec:method:jets}).
All EMDs are calculated with $\beta=2$, so the case of~\IsoRingTwo~is similar to the transverse thrust.
This choice of squared distances yields larger penalties for large displacements relative to small ones.
This is motivated by the study of jets rather than particles, to reduce differences between these two pictures of the event.
The squared distance measure is also computed more efficiently.
Every observable is separately normalised by the distance between the reference geometry (cylinder, ring, or dipole) and the maximally distant particle configuration with zero net transverse momentum such that the observable is defined on a range of $[0, 1]$.
These observables are summarised in Table~\ref{tab:intro:geometries}, along with their corresponding ground measures, following the implementations in Ref.~\cite{Cesarotti:2020hwb}.
The quasi-uniform reference geometries are illustrated in Figure~\ref{fig:intro:refgeos}, and a schematic illustration of the three event-shape observables studied in this analysis is provided in Figure~\ref{fig:geometry:carifigure2}.
 
\begin{table*}[htb]
\centering
\caption{
The different geometries used to define event isotropy, with their corresponding ground measures, and default quasi-uniform configurations, adapted from Ref.~\cite{Cesarotti:2020hwb} (where the dipole geometry was not considered explicitly).
Details of the normalisation of these observables can be found in Ref.~\cite{Cesarotti:2020hwb}.
}
\label{tab:intro:geometries}
\begin{tabular}{c @ {\quad} c@ {\quad} c}
\hline \hline
Geometry  & Ground Measure & $\mathcal{U}$\\
\hline
& & \\
Cylinder & $\theta^\text{cyl}_{ij} = \frac{12}{\pi^2 +16 y_\text{max}^2} \left( y_{ij}^2 + \phi_{ij}^2 \right) $ & $\mathcal{U}_{N}^\text{cyl}(|y|< y_{\text{max}})$ \\
& & \\
Ring & $\theta^\text{ring}_{ij} = \frac{\pi}{\pi -2} \left( 1 - \cos \phi_{ij}\right)$ & $\mathcal{U}_{N}^\text{ring}$\\
& & \\
Ring (Dipole) & $\theta^\text{ring}_{ij} = \frac{1}{1-\frac{1}{\sqrt{3}}} \left( 1 - \cos \phi_{ij}\right)$ & $\mathcal{U}_{2}^\text{ring}$\\
& & \\
\hline
\hline
& & \\
\end{tabular}
\end{table*}
 
Different ground measures $\theta_{ij}$ are chosen for the different reference geometries.
For the cylindrical case, the ground measure is taken to be the squared rapidity--azimuth distance between particles, where $y_i$ and $\phi_i$ are the rapidity and azimuth of each input particle, $y_{ij}$ and $\phi_{ij}$ are their differences between initial position $i$ and new position $j$, and $y_{\text{max}}=4.5$ is the maximum rapidity acceptance the cylinder.
For the ring geometry (both the isotropic and dipole configurations), the ground measure is taken to be the transversely projected opening angle $\phi$ between particles $i$ and $j$.
In all isotropy calculations, the reference geometry is oriented with respect to each event such that the overall EMD is minimised.
For large $N$, this is easily accomplished by azimuthally rotating the reference geometry such that a particle in the reference event is aligned with the leading jet in each event.
For the case of \IsoRingTwo, this minimisation is particularly important and the solution is non-trivial, akin to the computation of the thrust axis.
The relative azimuthal angle between $\mathcal{U}_{N=2}^\text{ring}$ and collider events that minimises the EMD is therefore found numerically using a function minimiser~\cite{2020SciPy-NMeth}.
 
\begin{figure}[htb]
\centering
\subfigure[]{\includegraphics[width=0.49\textwidth,trim=25 0 25 25,clip]{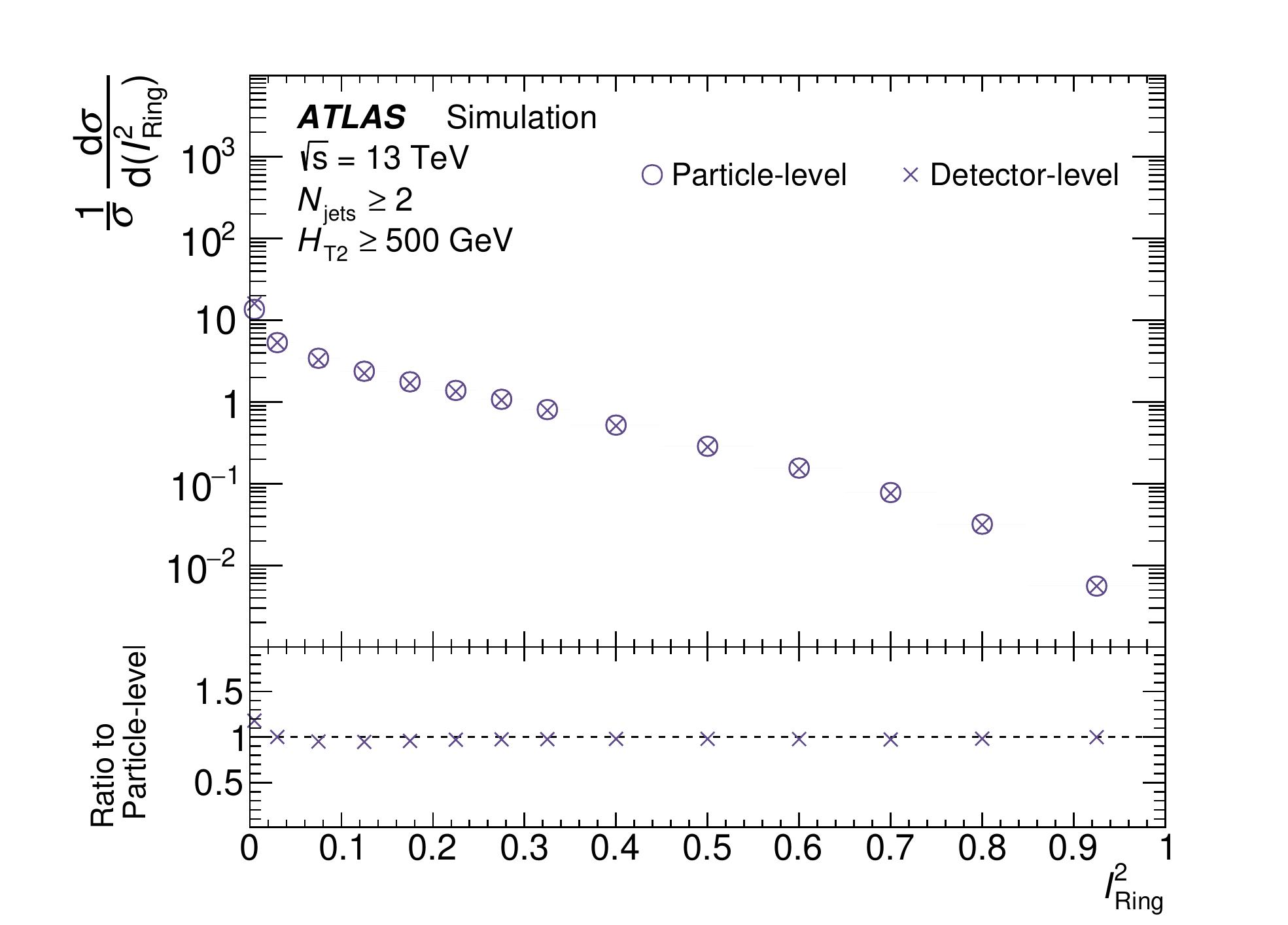}\label{fig:intro:detlevel:a}}
\subfigure[]{\includegraphics[width=0.49\textwidth,trim=25 0 25 25,clip]{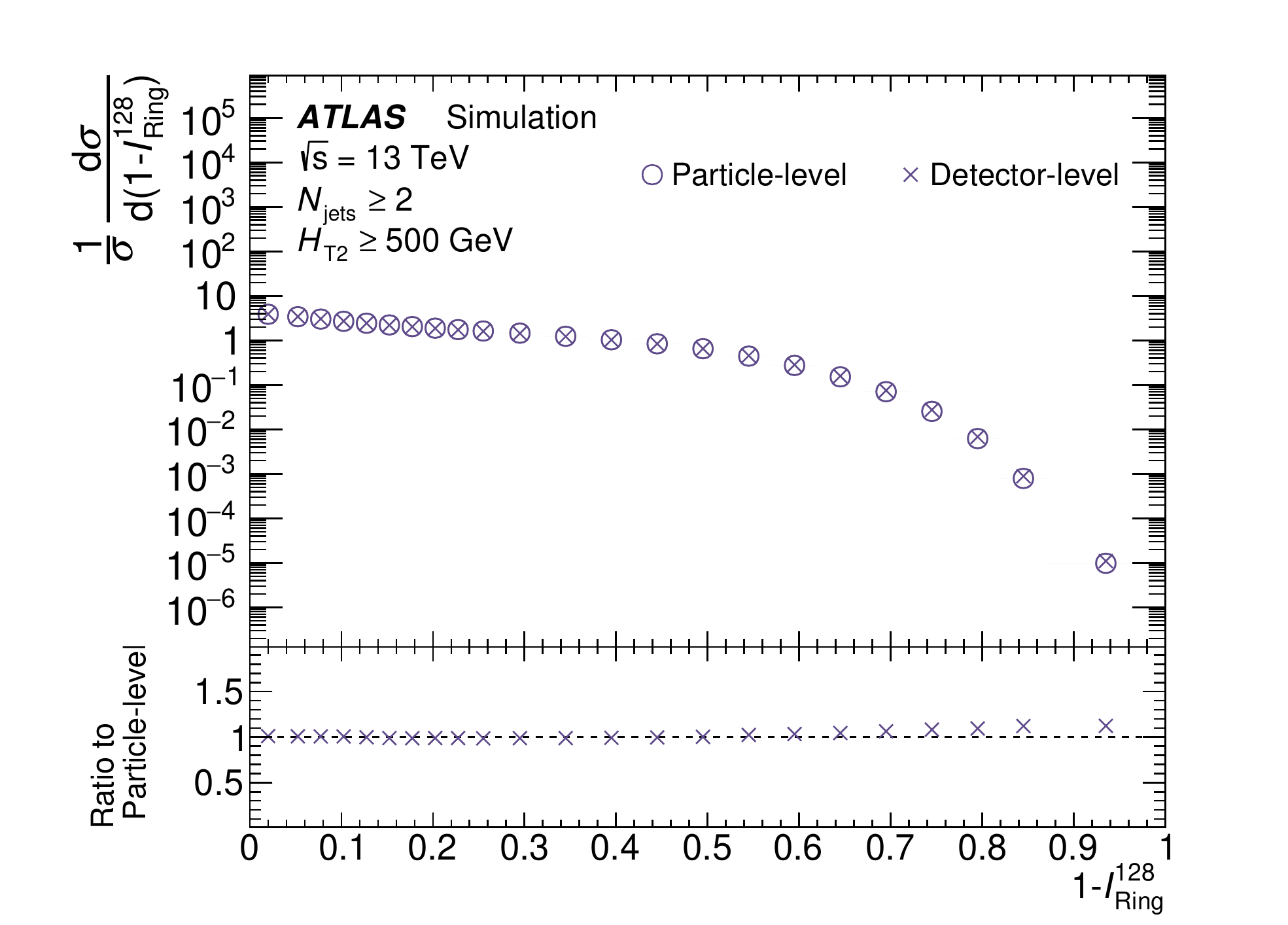}\label{fig:intro:detlevel:b}}\\
\subfigure[]{\includegraphics[width=0.49\textwidth,trim=25 0 25 25,clip]{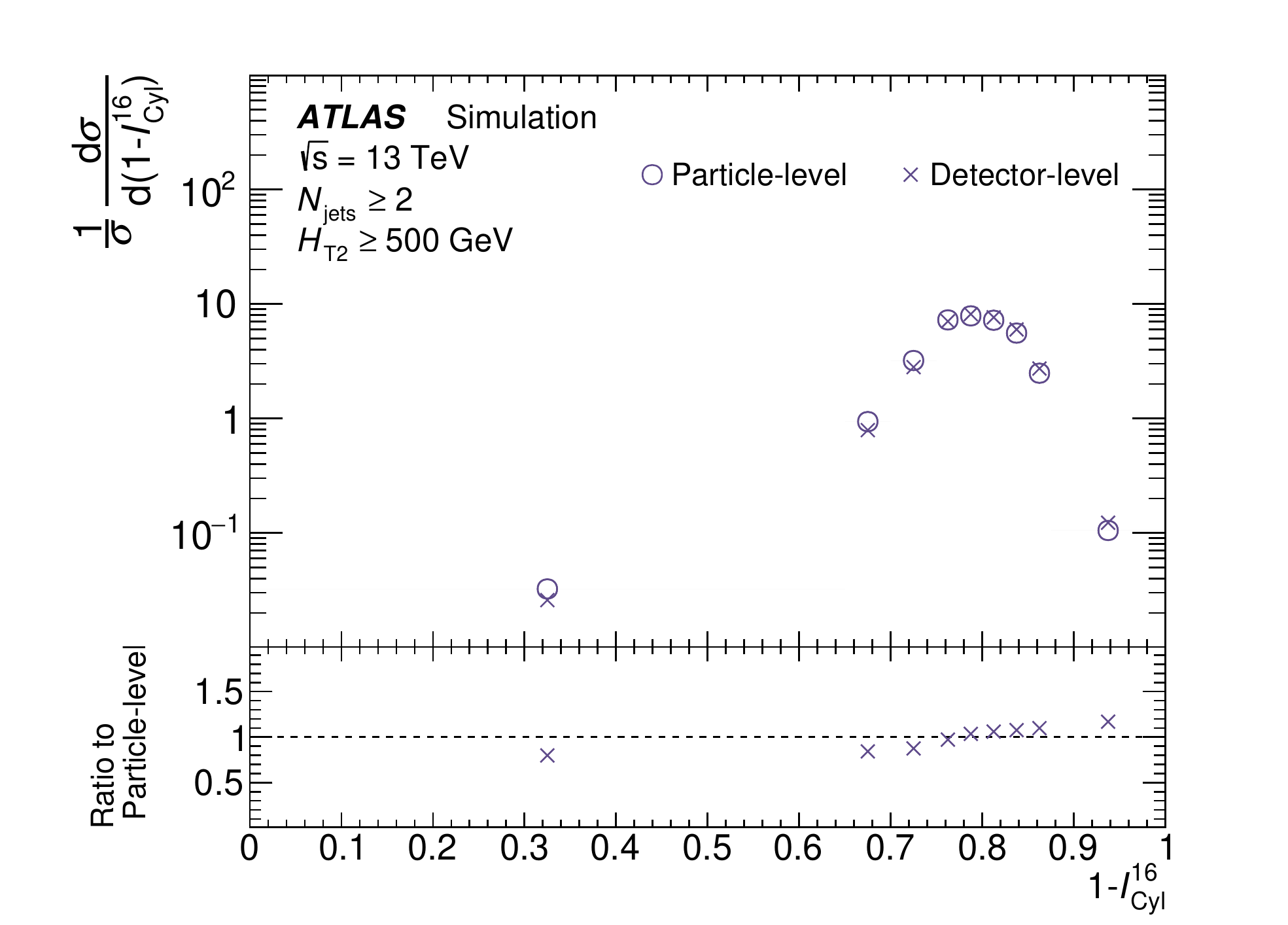}\label{fig:intro:detlevel:c}}
\caption{
Isotropy observables at particle level (open circles) and detector level (crosses), for events with $\HTTwo\geq500$~\GeV\ and $\njets\geq2$.
The event isotropy observables (a)~\IsoRingTwo, (b)~$1-\IsoRingOneTwentyEight$ and (c)~$1-\IsoCylSixteen$ are shown.
The lower panel of each figure displays the ratios of detector-level to particle-level distributions.
The plots are produced using simulated events generated with \PYTHIA[8.230].
}
\label{fig:intro:detlevel}
\end{figure}

To display the results of this analysis most clearly, all presented observables follow the historical convention that the least isotropic (`dijet-like') topology is near values of 0, and the most isotropic topology is near values of 1.
Therefore, the results of this measurement are presented in terms of $1-\IsoRingOneTwentyEight$, $1-\IsoCylSixteen$ and $\IsoRingTwo$.
Figure~\ref{fig:intro:detlevel} shows the distributions of these observables at particle level and detector level for the \PYTHIA sample of simulated events described in Section~\ref{sec:atlas:mc}, in events with $\njets\geq2$~and~$\HTTwo=(p_{\text{T},1}+p_{\text{T},2})\geq500~\GeV$ (Section~\ref{sec:methodology:selection}, for details about the binning see Section~\ref{sec:methodology:binning}).
For the ring-like geometries shown in Figures~\ref{fig:intro:detlevel:a}~and~\ref{fig:intro:detlevel:b}, it is clear that most multijet events passing this selection at the LHC are dijet-like and well-balanced, but significant tails extend into the isotropic regions.
The behaviour of $\IsoCylSixteen$, shown in Figure~\ref{fig:intro:detlevel:c}, is more complex and exhibits a `bulk' area with tails toward both isotropic topologies (large jet multiplicities, events with both central and forward jets) and non-isotropic topologies (jets only in one detector region, particularly the forward region on only one side of the event).
 
\begin{figure}[htb]
\centering
\subfigure[]{\includegraphics[width=0.49\textwidth]{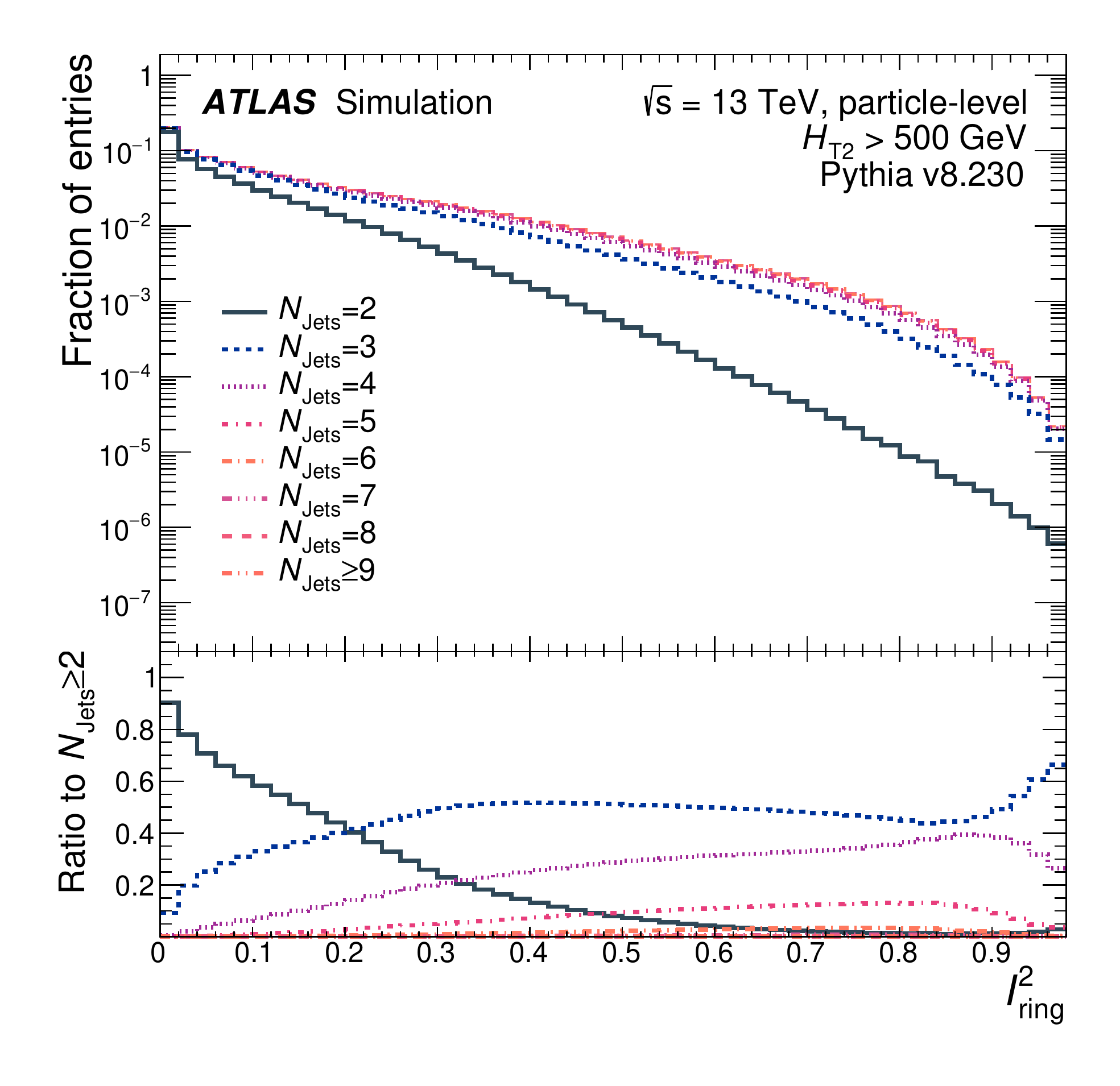}}
\subfigure[]{\includegraphics[width=0.49\textwidth]{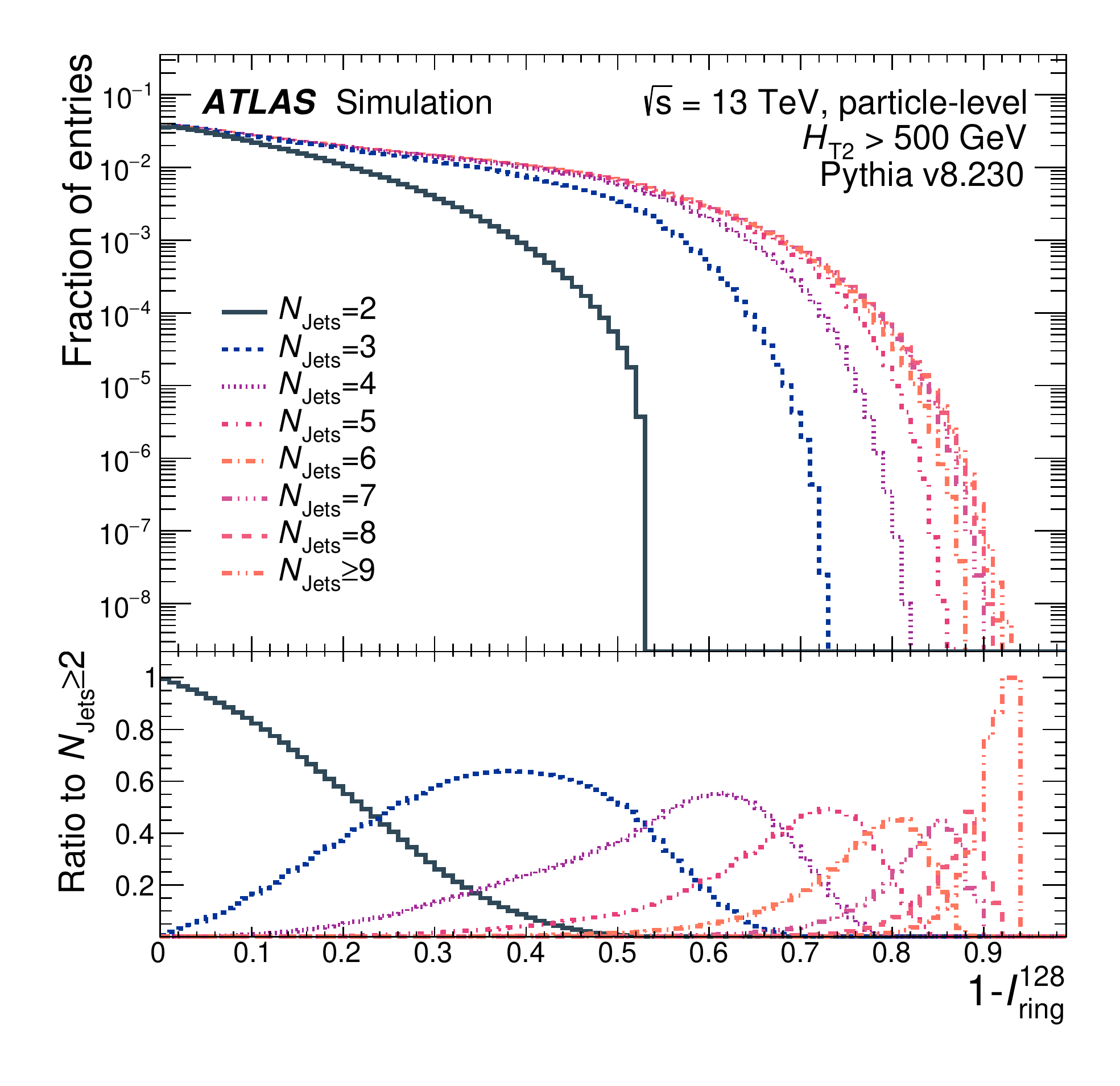}}
\caption{ Stacked histograms show the normalised (a) $\IsoRingTwo$ and (b) $1-\IsoRingOneTwentyEight$ distributions in separate bins of \njets, for particle-level \PYTHIA dijet events with $\njets\geq2$ and $\HTTwo > 500$~\GeV.
The lower panel of each figure shows the ratio of each \njets~bin to the inclusive distribution of the observable.
}
\label{fig:intro:diff}
\end{figure}

To better understand the differences between the shapes of $\IsoRingTwo$ and $1-\IsoRingOneTwentyEight$, it is informative to consider these observables in exclusive bins of jet multiplicity,~\njets.
These distributions are shown in Figure~\ref{fig:intro:diff}, demonstrating the improved performance of $1-\IsoRingOneTwentyEight$ relative to other event-shape observables in terms of selecting isotropic multijet events.
For $\IsoRingTwo$, each exclusive jet-multiplicity bin is distributed across the entire range of the event shape.
Three-jet events produce extremal values of $\IsoRingTwo$ most often, but such values may also be produced by events with other jet multiplicities.
The $1-\IsoRingOneTwentyEight$ observable has distinct endpoints for each exclusive jet-multiplicity bin, and the event shape distribution is dominated by events with increasing jet multiplicities as its value increases.
As jet multiplicities are sequentially ordered in this case, the higher cross-section for lower jet multiplicities results in a flatter distribution at lower values of $1-\IsoRingOneTwentyEight$ than for $\IsoRingTwo$, where the cross-section falls more steeply for values near 0.
 
\begin{figure}[htb]
\centering
\includegraphics[width=0.66\textwidth]{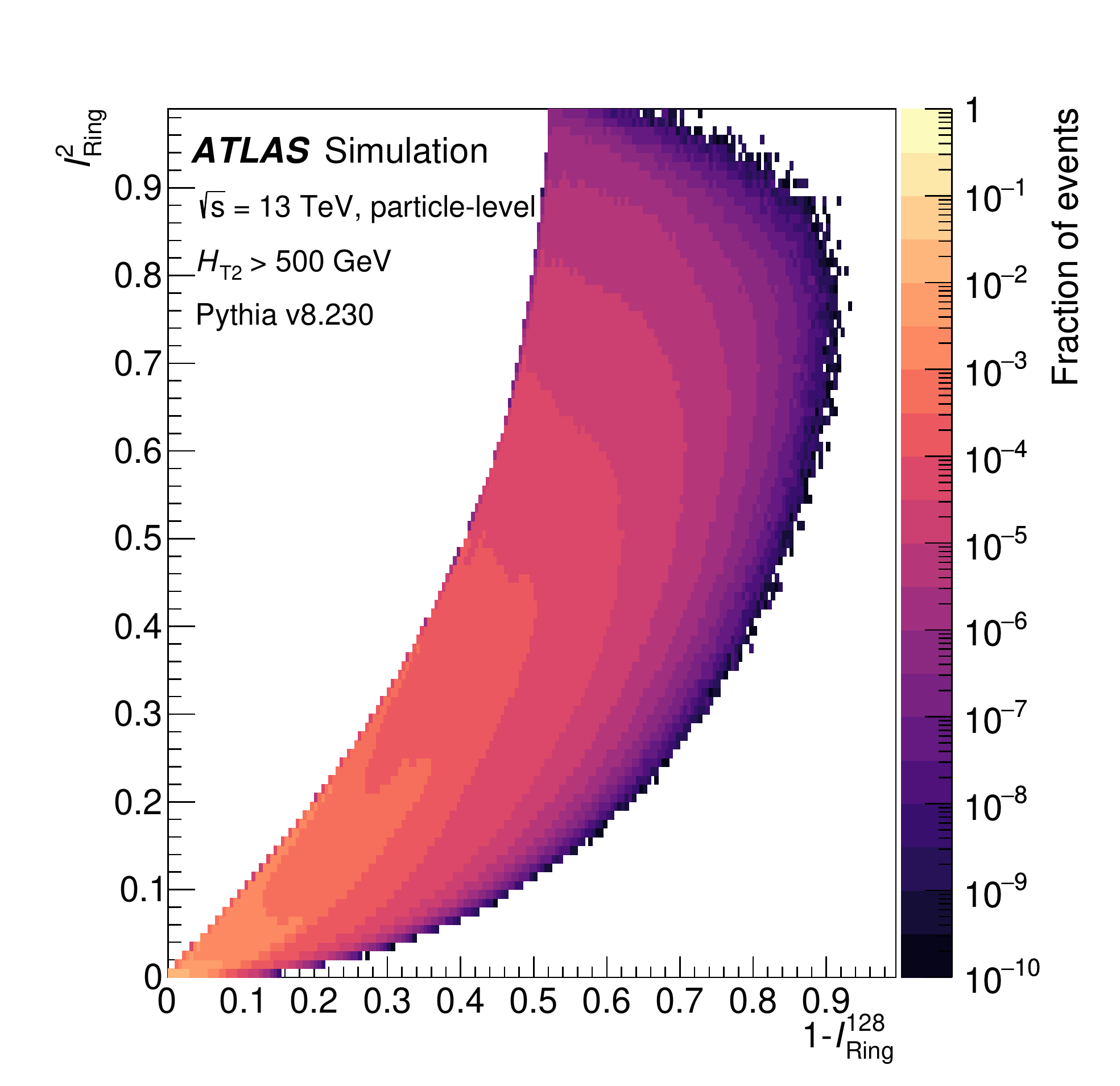}
\caption{ The correlation between $\IsoRingTwo$ and $1-\IsoRingOneTwentyEight$ is illustrated as a two-dimensional histogram, for events with $\njets\geq2$ and $\HTTwo > 500$~\GeV. }
\label{fig:intro:correlation}
\end{figure}

Even though there is a calculable EMD between the reference configurations themselves, the choice of ground measure used to define event isotropy observables results in a non-trivial relationship between $1-\IsoRingOneTwentyEight$~and~$\IsoRingTwo$.
This choice is also the reason that the two observables obtain their extreme values from different types of events.
The correlations between the studied event-isotropy observables are illustrated at particle level in Figure~\ref{fig:intro:correlation}.
This correspondence was discussed in Ref.~\cite{Cesarotti:2020hwb}, and results from the choice of ground measure and EMD angular exponent $\beta=2$, which causes the triangle inequality to be violated for these observables:
\begin{equation*}
0 \le \text{EMD}(\mathcal{E},\mathcal{E}^{\prime\prime}) \nless \text{EMD}(\mathcal{E},\mathcal{E}^{\prime}) + \text{EMD}(\mathcal{E}^\prime,\mathcal{E}^{\prime\prime}).
\end{equation*}
In fact, EMDs with this ground measure belong to a generalised class of ground measures known as $p$-Wasserstein metrics~\cite{wasserstein1969markov, dobrushin1970prescribing} (here, with $p=\beta$) that instead satisfy a generalised version of the triangle inequality:
\begin{equation*}
\text{EMD}(\mathcal{E}, \mathcal{E}^{\prime\prime})^{1/\beta} \le \text{EMD}(\mathcal{E},\mathcal{E}^{\prime})^{1/\beta} + \text{EMD}(\mathcal{E}^{\prime},\mathcal{E}^{\prime\prime})^{1/\beta}.
\end{equation*}
where the parameter $\beta$ is the exponent related to angular weighting in Equation~\ref{eq:emd} (taken here as $\beta=2$).
The non-trivial relationship between these observables motivates measurements of multiple reference-particle configurations with the same dimensionality -- i.e.\ measuring both $1-\IsoRingOneTwentyEight$~and~\IsoRingTwo.
 
\FloatBarrier
 
\section{The ATLAS detector, data and simulation}
\label{sec:atlas}

\newcommand{\AtlasCoordFootnote}{
ATLAS uses a right-handed coordinate system with its origin at the nominal interaction point (IP) in the centre of the detector and the \(z\)-axis along the beam pipe. The \(x\)-axis points from the IP to the centre of the LHC ring, and the \(y\)-axis points upwards. Cylindrical coordinates \((r,\phi)\) are used in the transverse plane, \(\phi\) being the azimuthal angle around the \(z\)-axis. The pseudorapidity is defined in terms of the polar angle \(\theta\) as \(\eta = -\ln \tan(\theta/2)\). Angular distance is measured in units of \(\Delta R \equiv \sqrt{(\Delta y)^{2} + (\Delta\phi)^{2}}\), where $y=(1/2)[(E+p_z)/(E-p_z)]$ is the object's rapidity defined by its energy and longitudinal momentum.}

\subsection{The ATLAS detector}
 
The ATLAS detector~\cite{PERF-2007-01} at the LHC covers nearly the entire solid angle around the collision point.\footnote{\AtlasCoordFootnote}
It consists of an inner tracking detector surrounded by a thin superconducting solenoid, electromagnetic and hadron calorimeters,
and a muon spectrometer incorporating three large superconducting air-core toroidal magnets.
 
The inner-detector system is immersed in a \SI{2}{\tesla} axial magnetic field
and provides charged-particle tracking in the range \(|\eta| < 2.5\).
The high-granularity silicon pixel detector covers the vertex region and typically provides four measurements per track,
the first hit normally being in the insertable B-layer 
installed before \RunTwo~\cite{ATLAS-TDR-19,PIX-2018-001}.
It is followed by the silicon microstrip tracker
, which usually provides eight measurements per track.
These silicon detectors are complemented by the transition radiation tracker (TRT),
which enables radially extended track reconstruction up to \(|\eta| = 2.0\).
The TRT also provides electron identification information
based on the fraction of hits (typically 30 in total) above a higher energy-deposit threshold corresponding to transition radiation.
 
The calorimeter system covers the pseudorapidity range \(|\eta| < 4.9\).
Within the region \(|\eta|< 3.2\), electromagnetic calorimetry is provided by barrel and
endcap high-granularity lead/liquid-argon (LAr) calorimeters,
with an additional thin LAr presampler covering \(|\eta| < 1.8\)
to correct for energy loss in material upstream of the calorimeters.
Hadron calorimetry is provided by the steel/scintillator-tile calorimeter,
segmented into three barrel structures within \(|\eta| < 1.7\), and two copper/LAr hadron endcap calorimeters.
The solid angle coverage is completed with forward copper/LAr and tungsten/LAr calorimeter modules
optimised for electromagnetic and hadronic energy measurements respectively.
 
The muon spectrometer 
comprises separate trigger and
high-precision tracking chambers measuring the deflection of muons in a magnetic field generated by the superconducting air-core toroidal magnets.
The field integral of the toroids ranges between \num{2.0} and \SI{6.0}{\tesla\metre}
across most of the detector.
Three layers of precision chambers, each consisting of layers of monitored drift tubes, cover the region \(|\eta| < 2.7\),
complemented by cathode-strip chambers in the forward region, where the background is highest.
The muon trigger system covers the range \(|\eta| < 2.4\) with resistive-plate chambers in the barrel, and thin-gap chambers in the endcap regions.
 
Interesting events are selected by the first-level trigger system implemented in custom hardware,
followed by selections made by algorithms implemented in software in the high-level trigger~\cite{TRIG-2016-01}.
The first-level trigger accepts events from the \SI{40}{\MHz} bunch crossings at a rate below \SI{100}{\kHz},
which the high-level trigger further reduces to record events to disk at about \SI{1}{\kHz}.
 
An extensive software suite~\cite{ATL-SOFT-PUB-2021-001} is used in data simulation, in the reconstruction and analysis of real
and simulated data, in detector operations, and in the trigger and data acquisition systems of the experiment.
 
\subsection{Data} 
 
This analysis is performed using data from LHC $pp$ collisions with $\sqrt{s}~=~13$~\TeV, collected during 2015--2018 with the ATLAS detector.
The total integrated luminosity of this dataset is $140~\ifb$.
The uncertainty in the combined 2015--2018 integrated luminosity is 0.83\% \cite{DAPR-2021-01}, obtained using the LUCID-2 detector \cite{LUCID2} for the primary luminosity measurements, complemented by measurements using the inner detector and calorimeters.
Due to the high instantaneous luminosity and the large total inelastic proton--proton ($pp$) cross section, there are, on average, 33.7 simultaneous (`\pileup') collisions in each bunch crossing.
Data are required to satisfy certain quality requirements~\cite{DAPR-2018-01} to be included in the analysis.
 
\FloatBarrier
\subsection{Simulation}\label{sec:atlas:mc}
 
Samples of Monte Carlo (MC) simulated dijet and multijet events are used in this analysis.
Since the jet production cross-section is much larger than the cross-section for electroweak processes, the dijet and multijet samples are sufficient to describe the data.
 
\PYTHIA[8.230]~\cite{Sjostrand:2014zea,Sjostrand:2006za} is used as the nominal MC generator for this analysis, and is also referred to here as the `nominal' simulation.
Samples of $\TwoToTwo$ dijet events were simulated using the A14 tune~\cite{ATL-PHYS-PUB-2014-021}, the Lund string hadronisation model and the \NNPDF[2.3lo]~\cite{Ball:2012cx} leading-order (LO) parton distribution function (PDF) set.
The \PYTHIA parton shower (PS) algorithm uses a dipole-style \pt-ordered evolution, and its renormalisation and factorisation scales were set to the geometric mean of the squared transverse masses of the outgoing particles.
\EVTGEN~\cite{Lange:2001uf} was used to model decays of heavy-flavour hadrons.
 
Two sets of \SHERPA[2.2.5]~\cite{Gleisberg:2008ta} dijet events were used with the default AHADIC cluster hadronisation model~\cite{Webber:1983if} or with the \SHERPA interface to the Lund string hadronisation model as implemented in \PYTHIA[6.4], and its decay tables. 
These samples include LO matrix element calculations for \TwoToTwo processes, and use the \SHERPA parton shower algorithm based on Catani--Seymour dipole subtraction~\cite{Schumann:2007mg}.
The \CT[14nnlo] next-to-next-to-leading-order (NNLO) PDF~\cite{Dulat:2015mca} set is used for matrix element calculations and \CT[10] is used for multi-parton interactions (MPI)~\cite{Lai:2010vv}.
 
Two sets of \HERWIG[7.1.3]~\cite{Bahr:2008pv,Bellm:2015jjp,Bellm:2017bvx} multijet events were generated with the \MMHT[nlo] PDF set~\cite{Harland-Lang:2014zoa}, default cluster hadronisation model and either the default angle-ordered PS or alternative dipole PS~\cite{Webber:1983if}.
These samples model \TwoToTwo~matrix elements with NLO accuracy and~\TwoToThree~matrix elements with LO accuracy.
Both parton shower models were matched to the matrix element calculation using the \MCatNLO matching scheme~\cite{Lonnblad:2001iq,Lonnblad:2011xx}.
The \pt of the leading jet is taken as the renormalisation scale.
 
Two additional samples of dijet events with NLO matrix element accuracy were produced with \POWHEG[v2]~\cite{Nason:2004rx,Frixione:2007vw,Alioli:2010xa} using the dijet process implemented in \POWHEGBOX[v2]~\cite{Alioli:2010xd}, matched to either the \PYTHIA[8] or angle-ordered \HERWIG[7] parton showers configured as for the corresponding samples described above.
The renormalisation and factorisation scales in these samples were set to the \pt of the underlying Born-level configuration.
For the \PYTHIA PS, the default Lund string hadronisation model was used with the \NNPDF[3.0nlo] PDF set~\cite{Ball:2014uwa} and A14 tune.
For the \HERWIG sample, the \NNPDF[3.0nlo] PDF set~\cite{Ball:2014uwa} was also used along with the default \HERWIG cluster-based hadronisation model.
These samples are referred to as the `\POWPY' and `\POWHER' samples.
 
All generated events were passed through a full detector simulation~\cite{SOFT-2010-01} based on \GEANT~\cite{Agostinelli:2002hh} and overlayed with simulated minimum-bias interactions generated using \PYTHIA[8]~with the A3 tune~\cite{ATL-PHYS-PUB-2016-017} and \NNPDF[2.3lo] PDF set~\cite{Ball:2012cx} to represent \pileup interactions.
The distribution of the average number of \pileup interactions in simulation is reweighted during data analysis to match that observed in \RunTwo data.
 
Additional details of the MC samples used in this measurement may be found in Ref.~\cite{ATL-PHYS-PUB-2019-017}.


\FloatBarrier
 
\section{Methodology}
\label{sec:methodology}

\subsection{Jets}\label{sec:method:jets}
 
All jets in this analysis are reconstructed using the \akt algorithm~\cite{Cacciari:2008gp} as implemented in \Fastjet~\cite{Fastjet}, using a jet radius parameter $R=0.4$.
The acceptance of jets at detector level has been increased relative to other recent event-shape measurements by ATLAS~\cite{STDM-2019-02}.
In particular, jets with lower transverse momentum and jets in the forward detector region have been included.
 
`Particle-level' jets are reconstructed in MC events without detector simulation. All detector-stable particles with a lifetime $\tau$ in the laboratory frame such that $c\tau > 10$~mm are used, except those particles that are expected to leave no or negligible energy depositions in the calorimeter, (i.e.\ neutrinos or muons).
Particle-level jets are required to have a $\pt > 60$~\GeV~and a rapidity $y$  satisfying $|y|<4.5$ to enter this analysis.
 
Detector-level jets are reconstructed from particle-flow objects~\cite{PERF-2015-09} that combine measurements from the ATLAS inner detector and calorimeter systems to improve the jet energy resolution (JER) and improve the jet reconstruction efficiency, especially at low jet \pt.
These jets are `cleaned' to remove those originating from detector noise, cosmic rays and beam-induced processes by following the methodology described in Ref.~\cite{ATLAS-CONF-2015-029}, updated for particle-flow jets but utilising the same observables.
In particular, the leading jet in each event is required to satisfy the `BadTight' jet cleaning criteria if it is within the inner detector acceptance ($|y|<2.4$).
Detector-level jets are required to have a $\pt > 60$~\GeV~and a rapidity $y$ satisfying $|y|<4.5$ to be retained for study.
 
The likelihood that a particle-flow jet originates from a \pileup interaction following these kinematic selections is sufficiently low that no additional \pileup jet rejection is applied~\cite{PERF-2014-03,PERF-2016-06}.
 
\subsection{Event selection}\label{sec:methodology:selection}
 
All detector-level events are required to have at least one vertex reconstructed from two or more inner-detector tracks with $\pt > 500$~\MeV, and to pass the data quality requirements described in Ref.~\cite{DAPR-2018-01}.
Events are required to have at least two selected jets ($\njets\geq2$) and to satisfy $\HTTwo \geq 400~\GeV$ to be included in the analysis.
Data were collected using a set of single-jet triggers~\cite{TRIG-2012-01}, whose thresholds varied depending on the data-taking period during \RunTwo.
The~\HTTwo~requirement is applied to ensure that the measurement is performed in a fiducial region where the single-jet triggers are fully efficient for the analysis selection.
 
Since the acceptance of the standard jet triggers decreases with increasing jet rapidity, they are combined with a dedicated set of forward-jet triggers.
Specific combinations of one central- and one forward-jet trigger are used to select events in ranges of~\HTTwo~where the combination is efficient.
Some triggers are prescaled during data-taking, so events in data are reweighted by the appropriate prescale factor to recover a smoothly falling jet \pt spectrum.
The prescale factors applied to central- and forward-jet triggers differ, so they are logically combined using the `inclusion method' of Ref.~\cite{Lendermann:2009ah}.

\subsection{Binning}\label{sec:methodology:binning}
 
In this analysis, the shape of the event isotropy observables \IsoCylSixteen, \IsoRingTwo~and~\IsoRingOneTwentyEight~is measured in inclusive bins of~\njets~and~\HTTwo.
The inclusive jet-multiplicity bins range from $\njets\geq2$ to $\njets\geq5$, and the inclusive bins of~\HTTwo~are~$\HTTwo\geq500$~\GeV, $\HTTwo\geq1000$~\GeV~and~$\HTTwo\geq1500$~\GeV.
Events with $\njets\geq2$~and~$\HTTwo \in [400, 500]~\GeV$ are included in the measurement only during the unfolding procedure (Section~\ref{sec:method:unfolding}), to mitigate the impact of migrations into the lowest~\HTTwo~bin of the measurement.
The detector resolution for events in the $\HTTwo \in [400, 500]~\GeV$ region was found to be worse than that for events in higher \HTTwo~bins, and so this bin was not included in measured region.
 
The final results (Section~\ref{sec:results}) are normalised such that their integral is equal to unity for each set of minimum $\njets$~and~$\HTTwo$ requirements applied.
Information about the relative normalisation of the various bins studied is thus lost, in exchange for a more precise measurement of the distribution shapes.
 
\FloatBarrier
\subsection{Unfolding}\label{sec:method:unfolding}
 
All data presented in Section~\ref{sec:results} are unfolded using an Iterative Bayesian Unfolding (IBU) procedure~\cite{DAgostini:1995} to remove effects arising from the finite efficiency, acceptance and resolution of the ATLAS detector.
This unfolding algorithm was implemented using the \textsc{RooUnfold}~\cite{Adye:2011gm} toolkit.
Four iterations of the unfolding procedure are used for all observables because this minimises the total uncertainty of the measurement.
Unfolding for the multi-differential measurements of event isotropy in inclusive bins of~\njets~and~\HTTwo~is performed simultaneously, to allow the unfolding procedure to account for migrations between all analysis bins.
 
\section{Systematic uncertainties}\label{sec:systematics}
 
Many sources of systematic uncertainty are accounted for in this analysis; they are described in the following Sections~\ref{sec:systematics:unfolding}--\ref{sec:systematics:other}.
For each systematic uncertainty, a varied response matrix is constructed and used in place of the nominal one during the unfolding procedure.

\subsection{Unfolding methodology: statistical uncertainties and non-closure}\label{sec:systematics:unfolding}
 
Statistical uncertainties arising from the finite Monte Carlo and data sample sizes used for this measurement are estimated during the unfolding procedure with Poissonian pseudo-experiments.
For the Monte Carlo simulation, pseudo-experiments are used to vary the response matrix used for the unfolding procedure.
The input MC prior is then unfolded with the varied response matrix; the efficiencies and acceptances are allowed to vary during this process.
For the data statistical uncertainty, pseudo-experiments are generated to vary the input data spectrum for the unfolding procedure and are then unfolded using the nominal \PYTHIA response matrix.
Five-hundred pseudo-experiments are generated in both cases; using larger numbers of pseudo-experiments does not significantly alter the results.
The 68\% inter-quantile range of the output distributions generated as a result of these variations is taken as the corresponding statistical uncertainty.
 
The non-closure uncertainty in the unfolding procedure is evaluated using a data-driven reweighting procedure.
The detector-level \PYTHIA spectrum is reweighted to match the observed data spectrum and then unfolded with the nominal \PYTHIA response matrix.
The difference between this unfolded spectrum and the nominal \PYTHIA particle-level spectrum is taken as a systematic uncertainty.
 
\subsection{Choice of nominal Monte Carlo generator}\label{sec:systematics:modelling}
 
In order to unfold a distribution, one relies on some nominal Monte Carlo simulation to construct the response matrix applied to data.
No particular MC model matches the data perfectly, so different results will be obtained if a different MC model is used in the unfolding procedure.
To account for the uncertainty related to the choice of nominal MC model, the unfolding procedure is repeated with the nominal \PYTHIA prior but using an alternative MC simulation for the event sample.
The alternative sample used to evaluate this uncertainty is the \HERWIG sample with an angle-ordered parton shower algorithm, which varies many aspects of the simulation with respect to the nominal \PYTHIA sample (Section~\ref{sec:atlas:mc}).
Despite the numerous differences between these two simulated samples, they provide competitive descriptions of the measured data, and the \HERWIG sample was also considered as a plausible choice for the nominal Monte Carlo model.
 
The effects of changing the MC model on the analysis efficiencies, acceptance and unfolding response matrix are considered individually.
An envelope of the observed differences between the final results following each of these three changes is constructed to conservatively estimate the uncertainty due to the choice of nominal MC generator.

\subsection{Jet energy scale and resolution}\label{sec:systematics:jesjer}
 
Systematic uncertainties in the $R=0.4$ jet energy scale (JES) and resolution (JER) are evaluated using a series of \emph{in situ} measurements and simulation-based techniques, thoroughly documented in Ref.~\cite{JETM-2018-05}.
The source of the largest single experimental uncertainty throughout the analysis is related to the jet energy resolution measurement, made using the \pT balance of dijet events.
 
Other relevant uncertainties arise from differences in the gluon-initiated jet energy response between \PYTHIA and \HERWIG (`jet-flavour response' in Ref.~\cite{JETM-2018-05}), and from the relative \emph{in situ} JES calibration.
For all JES/JER variations, the isotropy calculation is repeated with the varied set of jets.
 
The JES/JER uncertainties can potentially result in asymmetric variations, so they are left unsymmetrised in the presentation of the final results.
 
\subsection{Other experimental uncertainties}\label{sec:systematics:other}
 
Other uncertainties related to experimental effects are accounted for in this analysis.
They are typically small, but can occasionally be significant in certain measurement bins.
 
An uncertainty in the absolute luminosity measurement is applied as a $0.83\%$ variation to the normalisation of the nominal \PYTHIA MC simulation.
Due to the normalisation applied in this measurement, this systematic uncertainty cancels out by construction.
 
The uncertainty due to the mismodelling of pile-up events is negligible in all of the final results.
 
During certain \RunTwo data-taking periods, specific tile modules in the hadron calorimeter were disabled due to technical problems.
Some of these modules are also disabled in the simulated events corresponding to a given data-taking period, while other modules that were temporarily disabled during data-taking were not disabled in the simulation.
No additional correction is applied to the \pt of jets which may have deposited energy in disabled tile modules.
The impact of the disabled tile modules on the unfolded distributions is evaluated by repeating the measurement while vetoing events with jets directed at disabled modules in either data or the nominal \PYTHIA sample.
Differences between these `vetoed-event' results and the nominal set are taken as a source of systematic uncertainty.


\FloatBarrier
 
\section{Results}
\label{sec:results}

A representative selection of the measured distributions are presented in this section.
The systematic uncertainties are shown in a summarised format for clarity.
Uncertainties arising from similar sources are grouped as follows:
\begin{itemize}
\item \textbf{Stat.:} statistical uncertainties related to both data and MC sample size in the unfolding procedure (Section~\ref{sec:systematics:unfolding}).
\item \textbf{Unfolding:} the data-driven non-closure uncertainty in the unfolding procedure (Section~\ref{sec:systematics:unfolding}).
\item \textbf{MC model:} uncertainty related to the choice of MC models (Section~\ref{sec:systematics:modelling}), obtained by using the \HERWIG sample with angle-ordered parton showers rather than the nominal \PYTHIA MC sample when performing the unfolding procedure.
\item \textbf{JES+JER:} all uncertainties originating from the jet energy scale and jet energy resolution (Section~\ref{sec:systematics:jesjer}) -- the jet energy resolution uncertainty dominates this category in nearly all cases.
\item \textbf{Exp.\ conditions}: uncertainties related to the experimental conditions, such as those originating from \pileup reweighting and disabled tile modules (Section~\ref{sec:systematics:other}).
\end{itemize}
 
The unfolded data are compared with predictions from several state-of-the-art Monte Carlo models (Section~\ref{sec:atlas:mc}).
Good agreement is often observed between the leading-order and next-to-leading-order Monte Carlo generators throughout the non-isotropic region of a given distribution (i.e.\ for dijet-like events); poorer agreement is seen as particle configurations become more isotropic.
 
Figure~\ref{fig:results:incl:iring2} shows the most inclusive measurement of~\IsoRingTwo, in events with $\njets~\geq 2$ and $\HTTwo \geq 500~\GeV$.
Events with minimal values of this observable are balanced dijet events (e.g.\ Figure~\ref{fig:results:displays:iring2:low}), while events with maximal values are symmetric trijet systems (e.g.\ Figure~\ref{fig:results:displays:iring2:high}).
The NLO \POWPY and \POWHER predictions overestimate the cross-section at intermediate values of~\IsoRingTwo, and underestimate the cross-section at low values.
The NLO \HERWIG predictions with angle-ordered parton showers are closest to the data for small values; their agreement is slightly poorer for isotropic events.
The \HERWIG sample with the dipole PS model appears to slightly overestimate the cross-section of extremely well-balanced events, but agrees with the angle-ordered model throughout the rest of the distribution.
Overall, the data are best described in the isotropic region by the MC predictions with NLO matrix element calculations.
Leading-order \PYTHIA and \SHERPA predictions describe the back-to-back and intermediate range of the distribution well, but underestimate the cross-section for the most isotropic events (for values above $\sim0.6$).
No significant differences are observed between the cluster and Lund string hadronisation models for the \SHERPA samples.
The dominant systematic uncertainties of the measured distribution are related either to the jet energy resolution or to the choice of MC model used in the unfolding for isotropic events.
The total uncertainty of the measured distribution is below 5\% except in the most isotropic bin, where the uncertainty due to the choice of MC model becomes large.
 
\begin{figure}[htb]
\centering
\includegraphics[width=0.70\textwidth,trim=25 0 25 25,clip]{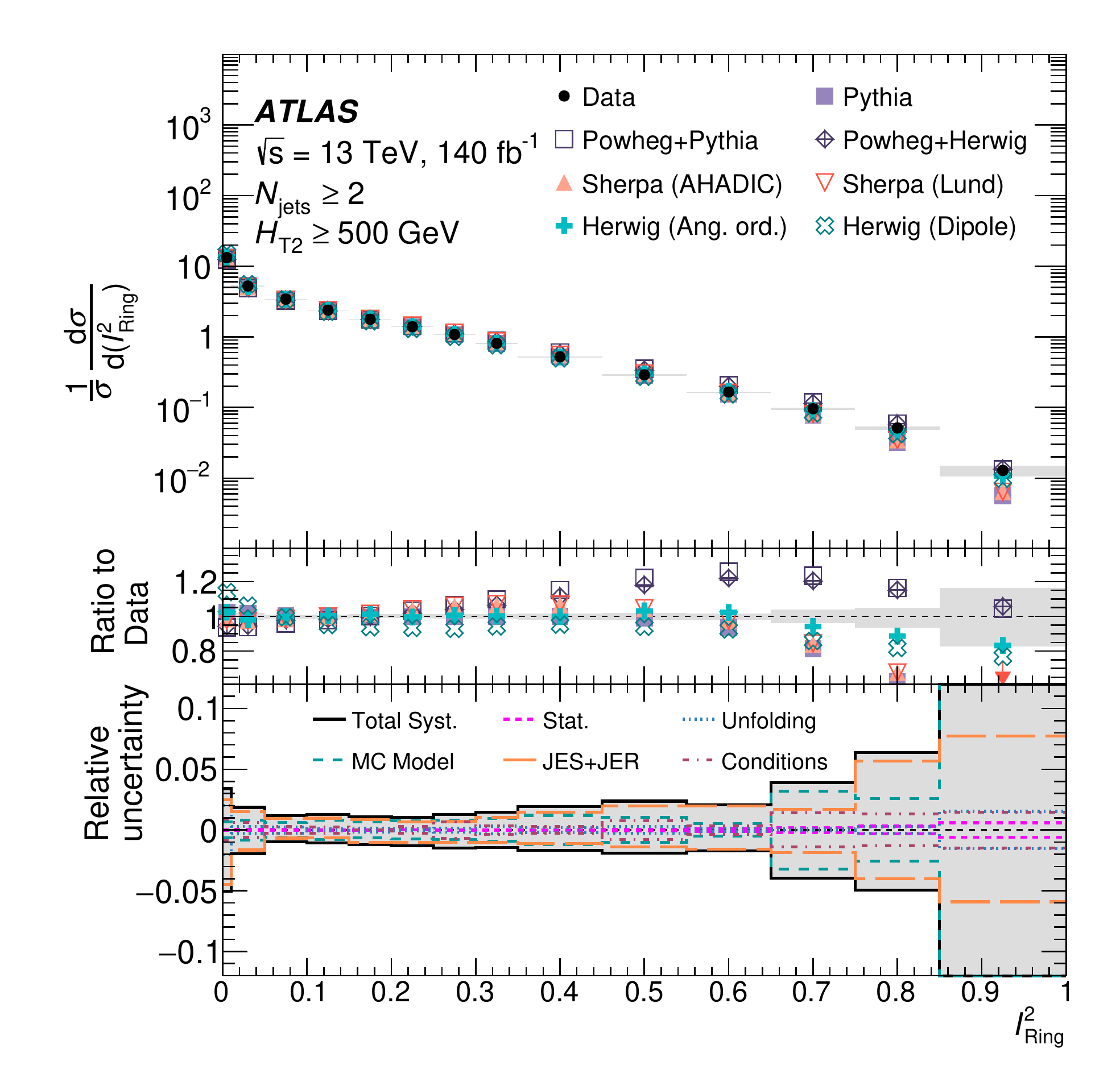}
\caption{The shape-normalized~\IsoRingTwo~cross-section in data (closed circles), compared with predictions from several Monte Carlo generators.
Events with $\HTTwo\geq500~\GeV$ and $\njets\geq2$ are included.
The middle panel displays the ratios of different event generator predictions to the unfolded data.
Event generator predictions are displayed as different marker styles.
The grey band in the upper and middle panels indicates the total uncertainty of the measurement.
If the ratio of a prediction to the unfolded data is outside the range of values displayed in the middle panel, an arrow is drawn at the edge of the panel as an indicator.
The lower panel summarises the various sources of systematic uncertainty in the measurement.
Systematic uncertainties are summarized in groups, with different line styles.
The total uncertainty is shown as a solid black line.
Some uncertainty bands take values outside the range displayed in the figure: the range is selected for maximum clarity in the bulk of the distribution.
}
\label{fig:results:incl:iring2}
\end{figure}
 
\begin{figure}[htb]
\centering
\subfigure[]{\includegraphics[width=0.55\textwidth]{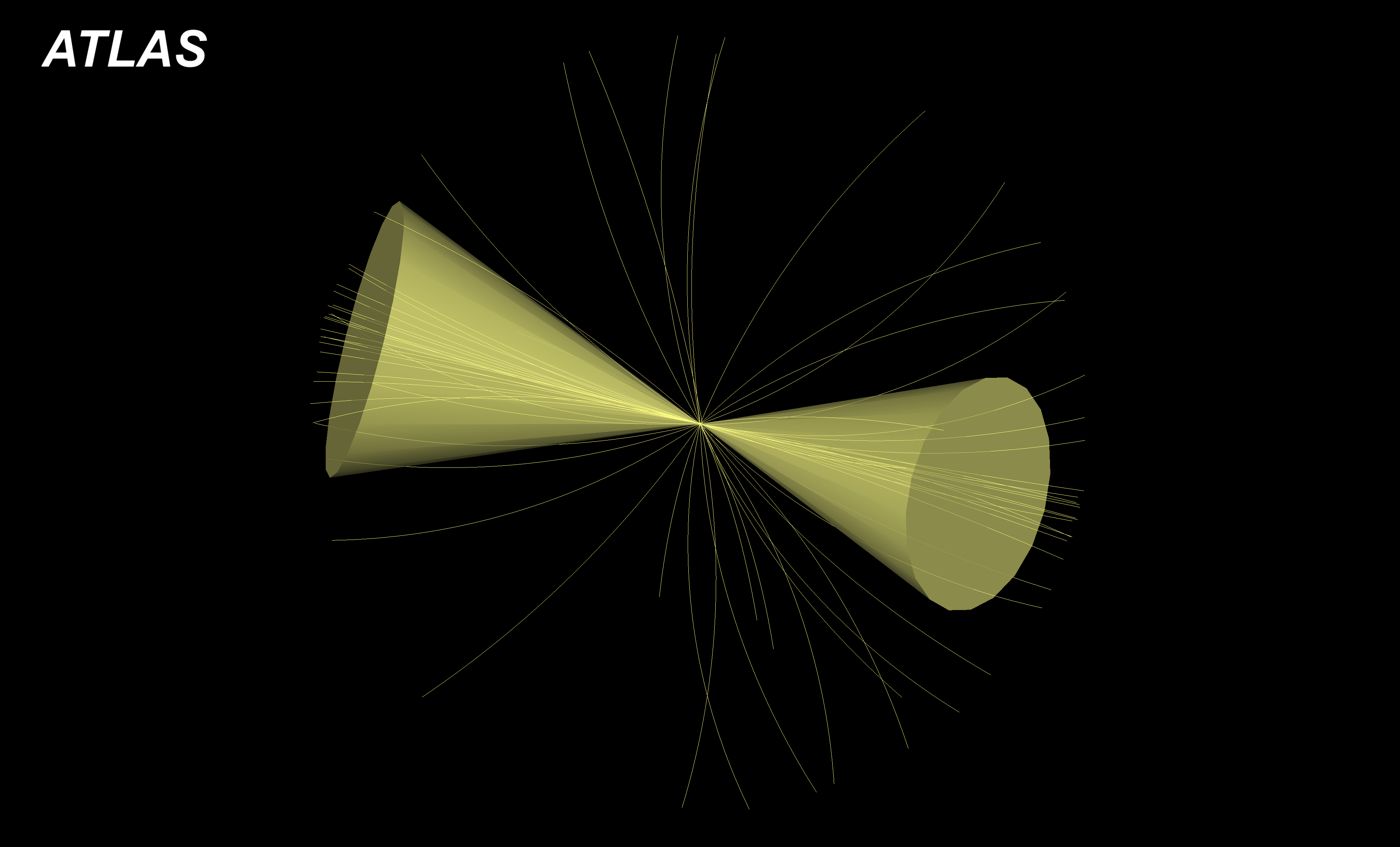}\label{fig:results:displays:iring2:low}}\\
\subfigure[]{\includegraphics[width=0.55\textwidth]{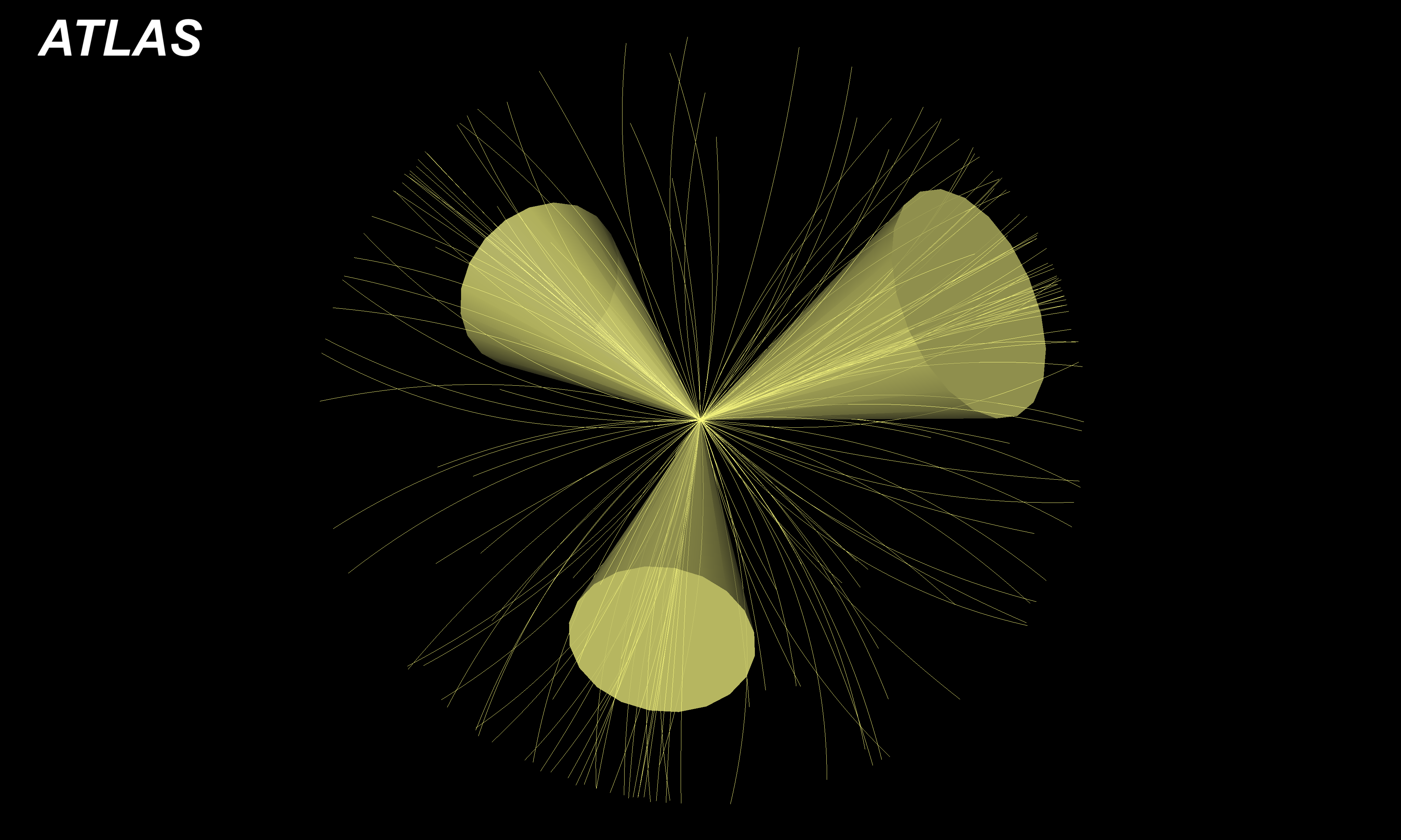}\label{fig:results:displays:iring2:high}}\\
\subfigure[]{\includegraphics[width=0.55\textwidth]{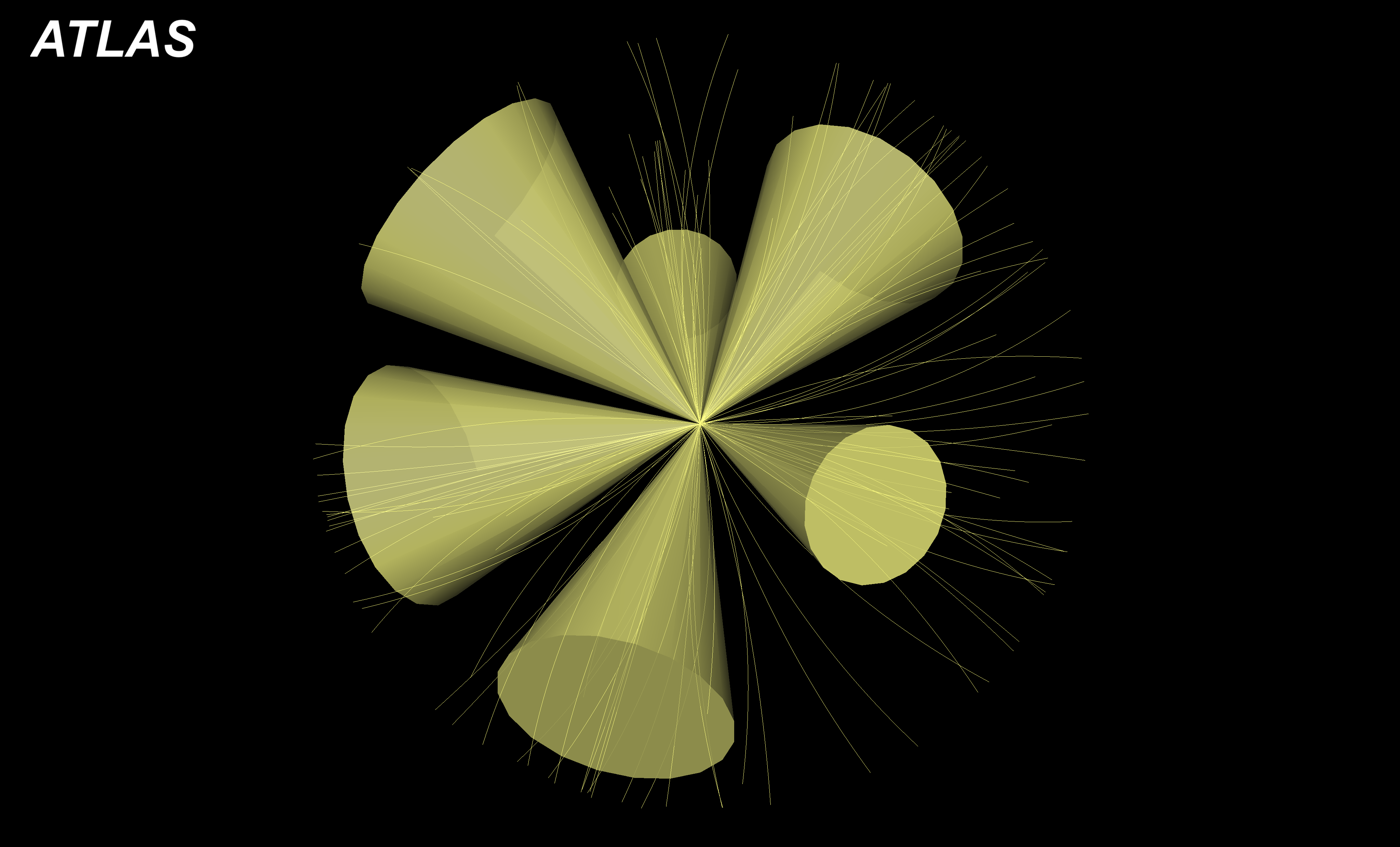}\label{fig:results:displays:iring128:high}}
\caption{
Displays of three events in the \RunTwo dataset that are examples of extreme values of the ring-like event-isotropy observables studied in this analysis.
The event displays show an image of the event in the transverse plane, with the beamline running perpendicularly into the images at their centres.
\Akt $R=0.4$ particle-flow jets passing a \pt requirement of 60~\GeV~are illustrated as cones in these displays, with a length corresponding to their logarithmically rescaled \pt.
Charged-particle tracks in the inner detector are also shown, as curved lines.
The events are (a) event 1921189174 from run 349268, which has small values of both \IsoRingTwo\ and $1-\IsoRingOneTwentyEight$, (b) event 1126942872 from run 305811, which has a large value of \IsoRingTwo\ and a moderate value of $1-\IsoRingOneTwentyEight$, and (c) event 2132056011 from run 349268, which has a large value of $1-\IsoRingOneTwentyEight$ and a moderate value of \IsoRingTwo.
}
\end{figure}

The $1-\IsoRingOneTwentyEight$~distribution is shown in Figure~\ref{fig:results:incl:iring128}, also for events with $\njets~\geq 2$ and $\HTTwo \geq 500~\GeV$.
Balanced dijet events (e.g.\ Figure~\ref{fig:results:displays:iring2:low}) produce the smallest values of this observable, while multijet events with isotropic energy arrangements (e.g.\ Figure~\ref{fig:results:displays:iring128:high}) produce the largest values.
The increased dynamic range of this observable is evident, as the measured cross-section spans approximately six orders of magnitude.
The quality of the modelling description for this observable differs from that of $\IsoRingTwo$~due to the different isotropic patterns it selects.
In particular, the \POWPY and \POWHER predictions are found to strongly disagree with those of the other MC generators, overestimating the measured cross-section for isotropic events while all other predictions underestimate it.
Large differences are also found between the \HERWIG angle-ordered and dipole shower models: the dipole model predicts relatively more dijet-like events than the angle-ordered model, and correspondingly fewer isotropic events.
No notable differences are seen between the \SHERPA hadronisation models, which together are found to come the closest to describing the measured data for larger values of $1-\IsoRingOneTwentyEight$.
The JES/JER systematic uncertainties are the most relevant source of uncertainty for most of the unfolded distribution, occasionally matched by the uncertainty related to the choice of nominal MC model for the unfolding procedure.
For the most isotropic events, statistical uncertainties become non-negligible, and the systematic uncertainty related to the effect of the hadron calorimeter's disabled tile modules during \RunTwo data-taking also becomes large.
The total uncertainty of this distribution is under 5\% until $1-\IsoRingOneTwentyEight \sim 0.6$, where it grows to be larger than 10\%--15\% for the most isotropic events.
 
\begin{figure}[htb]
\centering
\includegraphics[width=0.70\textwidth,trim=25 0 25 25,clip]{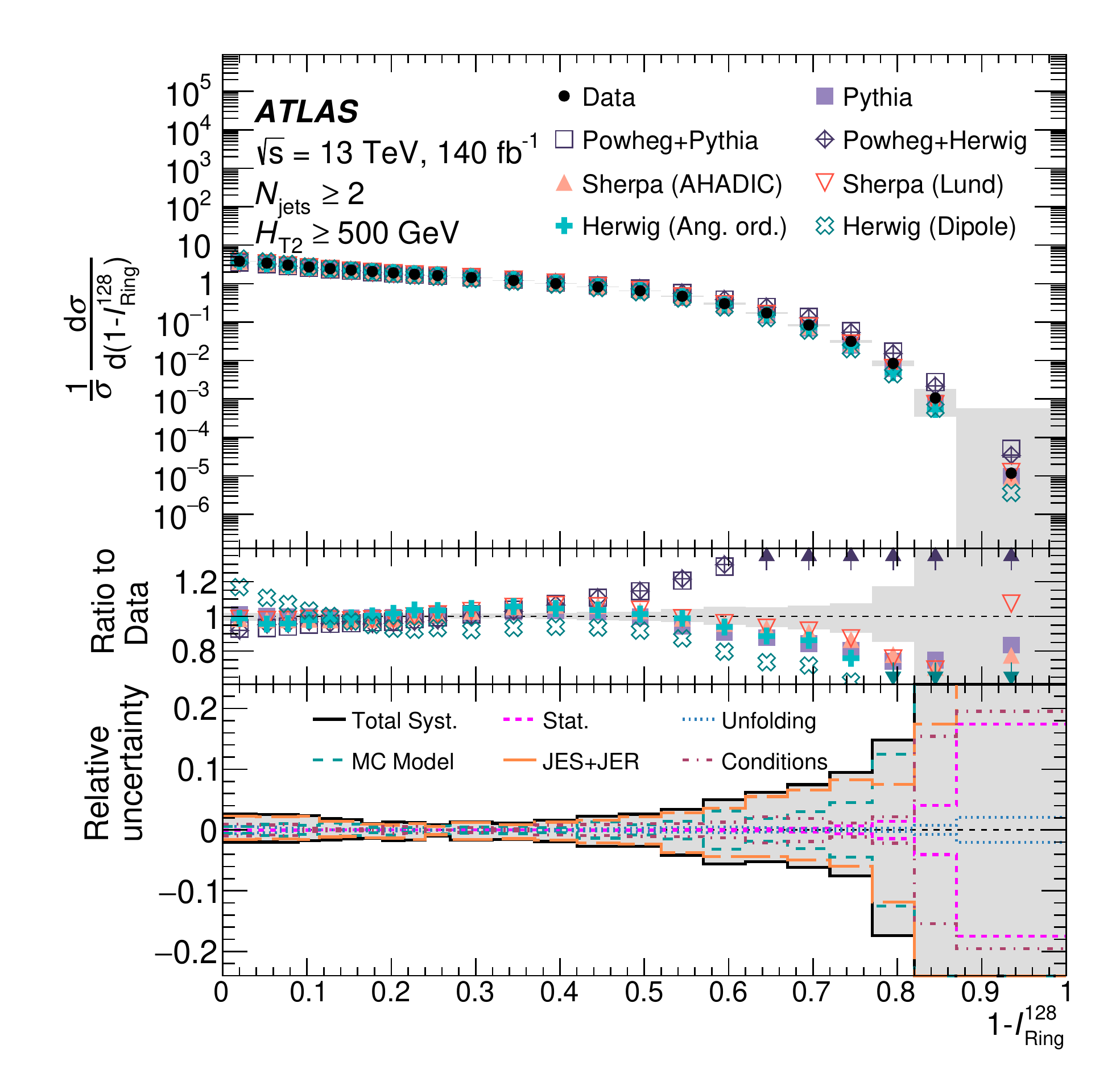}
\caption{The shape-normalized ~\IsoRingOneTwentyEight~cross-section in data (closed circles), compared with predictions from several Monte Carlo generators.
Events with $\HTTwo\geq500~\GeV$ and $\njets\geq2$ are included.
The middle panel displays the ratios of different event generator predictions to the unfolded data.
Event generator predictions are displayed as different marker styles.
The grey band in the upper and middle panels indicates the total uncertainty of the measurement.
If the ratio of a prediction to the unfolded data is outside the range of values displayed in the middle panel, an arrow is drawn at the edge of the panel as an indicator.
The lower panel summarises the various sources of systematic uncertainty in the measurement.
Systematic uncertainties are summarized in groups, with different line styles.
The total uncertainty is shown as a solid black line.
Some uncertainty bands take values outside the range displayed in the figure: the range is selected for maximum clarity in the bulk of the distribution.
}\label{fig:results:incl:iring128}
\end{figure}

The most inclusive measurement of the two-dimensional isotropy observable $1-\IsoCylSixteen$, for events with $\njets~\geq 2$ and $\HTTwo \geq 500~\GeV$, is shown in Figure~\ref{fig:results:incl:icyl16}.
This distribution exhibits different characteristics than the ring-like geometries. Events with dijet systems in the forward region on one side of ATLAS (e.g.\ Figure~\ref{fig:results:displays:icyl16:low}) produce the smallest values of this observable; the highest values are produced by multijet events that evenly cover the rapidity--azimuth plane with activity in both the central and forward regions (e.g.\ Figure~\ref{fig:results:displays:icyl16:high}).
None of the MC predictions accurately describe this observable, although the best descriptions occur near the peak of the distribution around $1-\IsoCylSixteen\sim0.8$.
The \HERWIG angle-ordered and dipole parton shower models predict distributions that have a peak at respectively larger and smaller values than that observed in the measured data.
This results in large differences between their respective ratios to the unfolded data.
As a result, they surround the data points across the entire distribution except for the highest-value bin.
The predictions from the \PYTHIA, \POWPY and \POWHER samples are consistent except at low values, where the \PYTHIA sample overestimates the observed cross-section.
Once again, no sensitivity to the hadronisation models implemented in \SHERPA is observed.
The precision of the measurement in this $\njets$ and $\HTTwo$ bin is everywhere better than 10\%, and is dominated throughout by the jet energy resolution component of the JES/JER error band.
 
\begin{figure}[htb]
\centering
\includegraphics[width=0.70\textwidth,trim=25 0 25 25,clip]{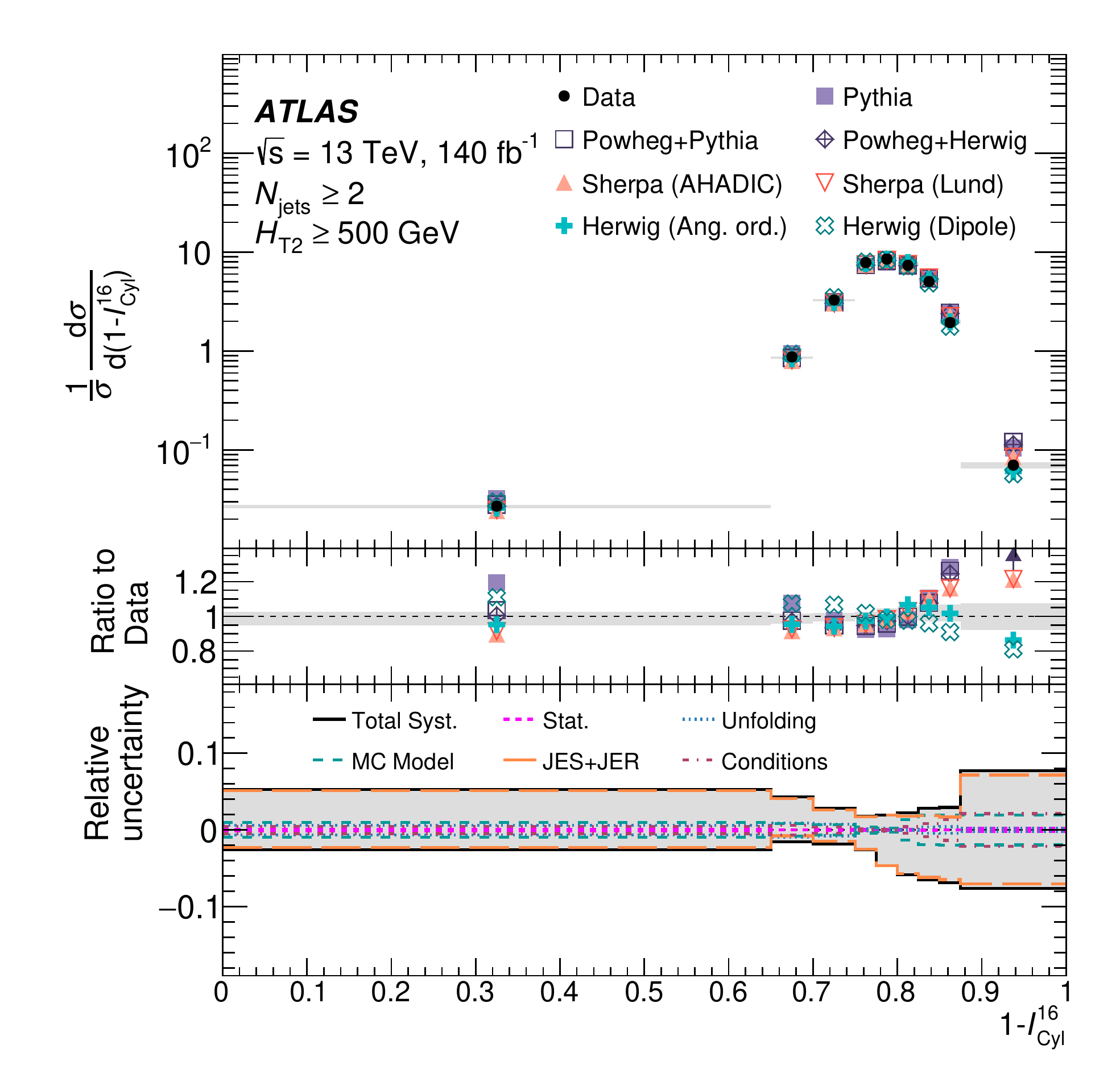}
\caption{The shape-normalized~\IsoCylSixteen~cross-section in data (closed circles), compared with predictions from several Monte Carlo generators.
Events with $\HTTwo\geq500~\GeV$ and $\njets\geq2$ are included.
The middle panel displays the ratios of different event generator predictions to the unfolded data.
Event generator predictions are displayed as different marker styles.
The grey band in the upper and middle panels indicates the total uncertainty of the measurement.
If the ratio of a prediction to the unfolded data is outside the range of values displayed in the middle panel, an arrow is drawn at the edge of the panel as an indicator.
The lower panel summarises the various sources of systematic uncertainty in the measurement.
Systematic uncertainties are summarized in groups, with different line styles.
The total uncertainty is shown as a solid black line.
}\label{fig:results:incl:icyl16}
\end{figure}
 
\begin{figure}[htb]
\centering
\subfigure[]{\includegraphics[width=0.60\textwidth]{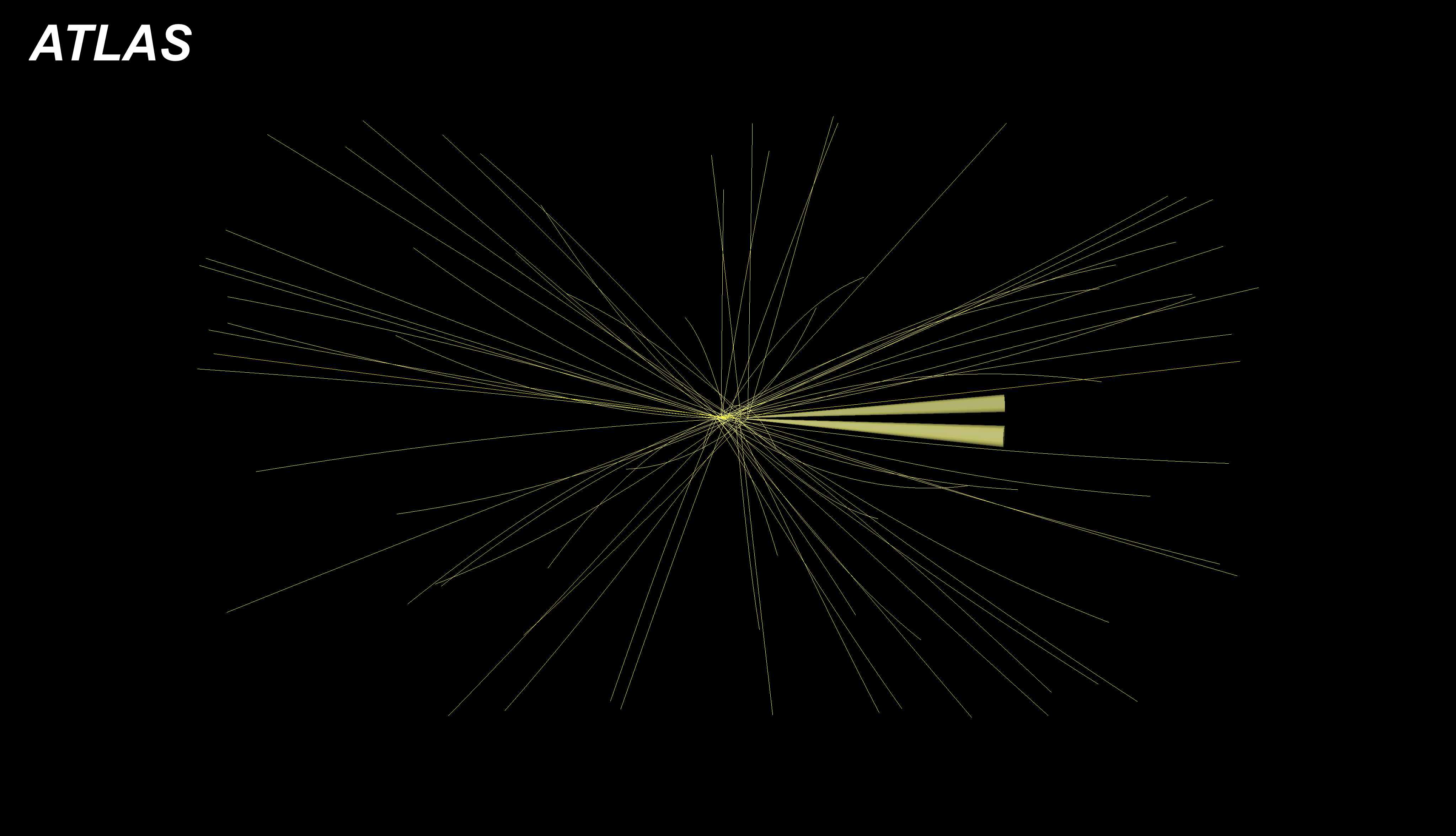}\label{fig:results:displays:icyl16:low}}\\
\subfigure[]{\includegraphics[width=0.60\textwidth]{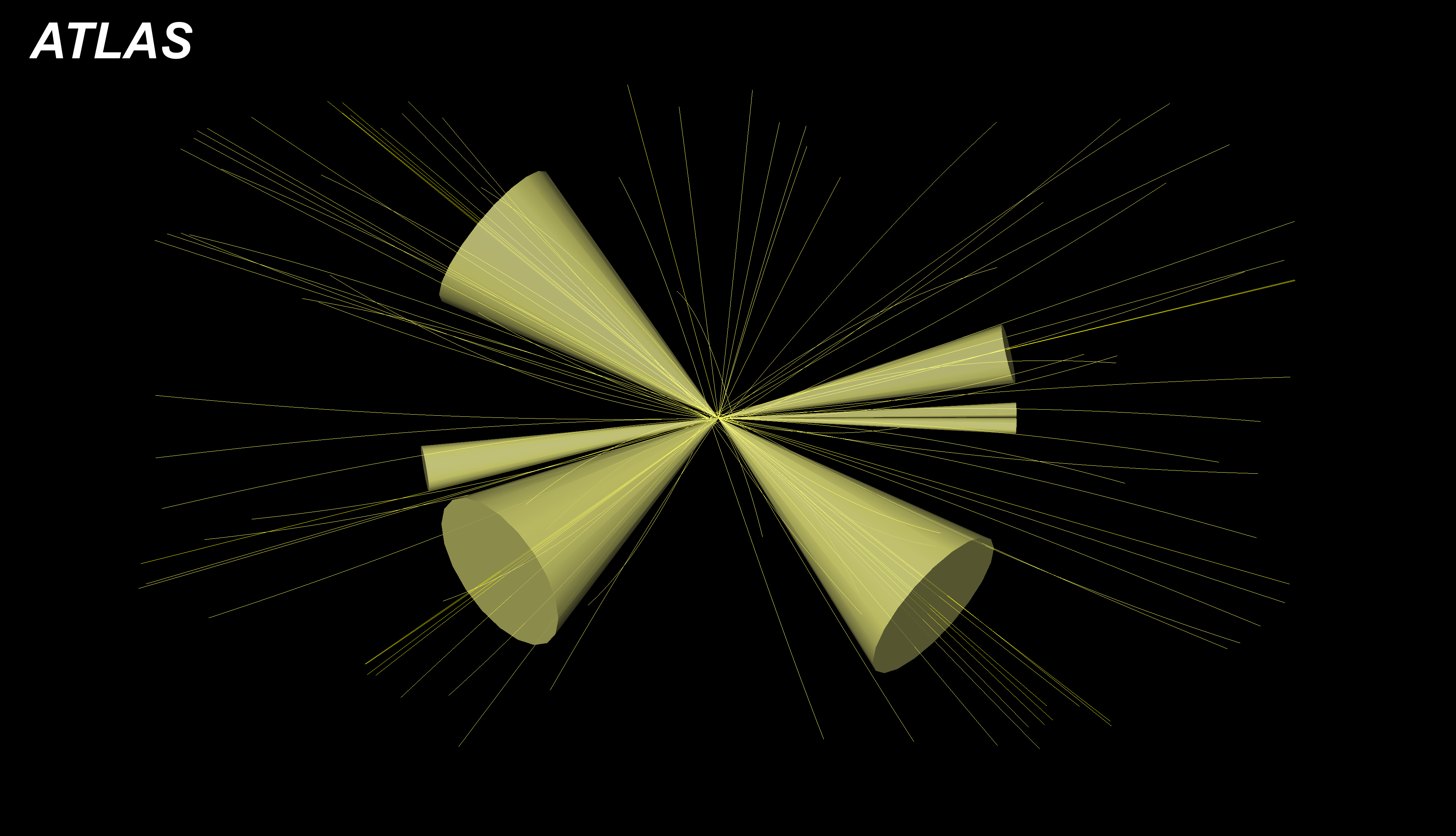}\label{fig:results:displays:icyl16:high}}\\
\caption{
Displays of two events in the \RunTwo dataset that are examples of extreme values of the cylindrical event-isotropy observables studied in this analysis.
The event displays show an image of the event from the side of the barrel, with the beamline running horizontally across the middle of each image.
\Akt $R=0.4$ particle-flow jets passing a \pt requirement of 60~\GeV~are illustrated as cones in these displays, with a length corresponding to their logarithmically rescaled \pt.
Charged-particle tracks in the inner detector are also shown, as curved lines.
The events are (a) event 1095666999 from run 307454, which has a small value of $1-\IsoCylSixteen$, and (b) event 2433141809 from run 340030, which has a large value of $1-\IsoCylSixteen$.
}
\end{figure}

Figures~\ref{fig:results:ison:iring2}--\ref{fig:results:ison:icyl16} present measurements of event isotropy observables with $\HTTwo\geq500~\GeV$ and an increasing minimum \njets~requirement.
Intuitively, the average value of each observable becomes larger as the minimum jet multiplicity is increased, indicating a more isotropic topology.
Binning in \njets~can elicit larger differences between the MC predictions, particularly between the angle-ordered and dipole \HERWIG parton shower models and the \PYTHIA, \POWPY and \POWHER predictions for back-to-back events.
Even at larger minimum jet multiplicities, the NLO MC predictions are found to maintain the quality of their description of the rate of balanced trijet events at large values of $\IsoRingTwo$.
The \HERWIG sample with the dipole PS model is observed to increasingly underestimate the cross-section of back-to-back events in the $\IsoRingTwo$ distribution as the jet multiplicity increases, while the angle-ordered PS model instead overestimates this region.
The differences between these models for the $1-\IsoCylSixteen$ distributions are also enhanced by increasing the minimum jet multiplicity.
 
The largest uncertainties in these measured distributions are again typically due to the JES/JER systematic uncertainties, although changing the MC model used in the unfolding procedure can result in larger uncertainties for larger jet-multiplicity values (and so, for larger values of $1-\IsoRingOneTwentyEight$).
In the tails of distributions, the statistical uncertainties and those related to disabled tile modules can become sizeable, but never dominant for the $\HTTwo\geq500~\GeV$ bin.
Overall, each observable is measured less precisely as the minimum~\njets~requirement is increased.
The \IsoRingTwo~distributions tend to be measured with a precision better than 10\%, except in the lowest and highest bins.
For $1-\IsoRingOneTwentyEight$, the uncertainty for low values is less than 5\% for $\njets\geq2, 3$ but increases in this region for larger jet multiplicities.
In the $\njets\geq5$ selection, the uncertainty in this region approaches 50\%.
 
\begin{figure}[htbp]
\centering
\subfigure[$\njets\geq2$]{\includegraphics[width=0.45\textwidth,trim=25 0 25 25,clip]{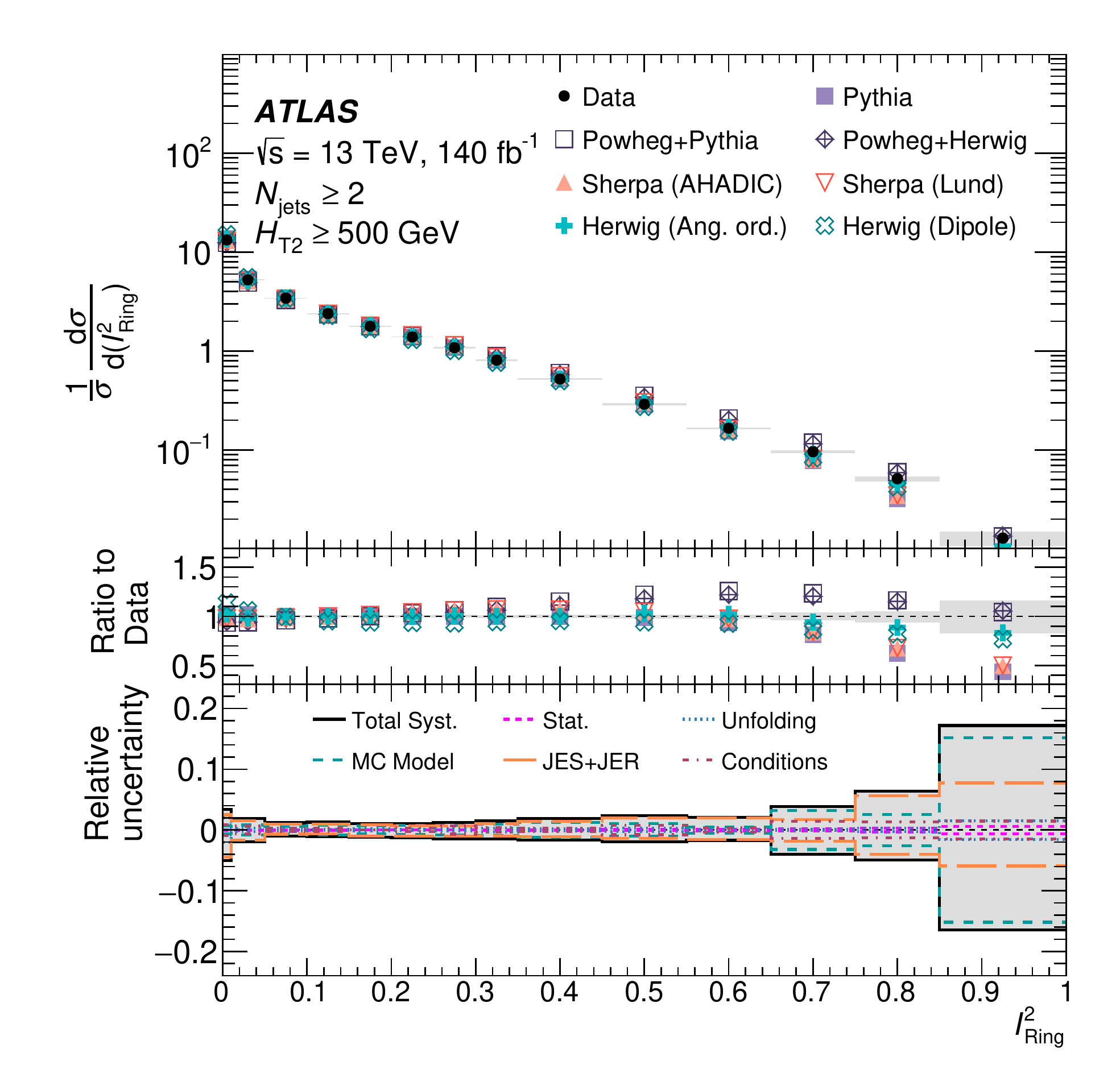}}
\subfigure[$\njets\geq3$]{\includegraphics[width=0.45\textwidth,trim=25 0 25 25,clip]{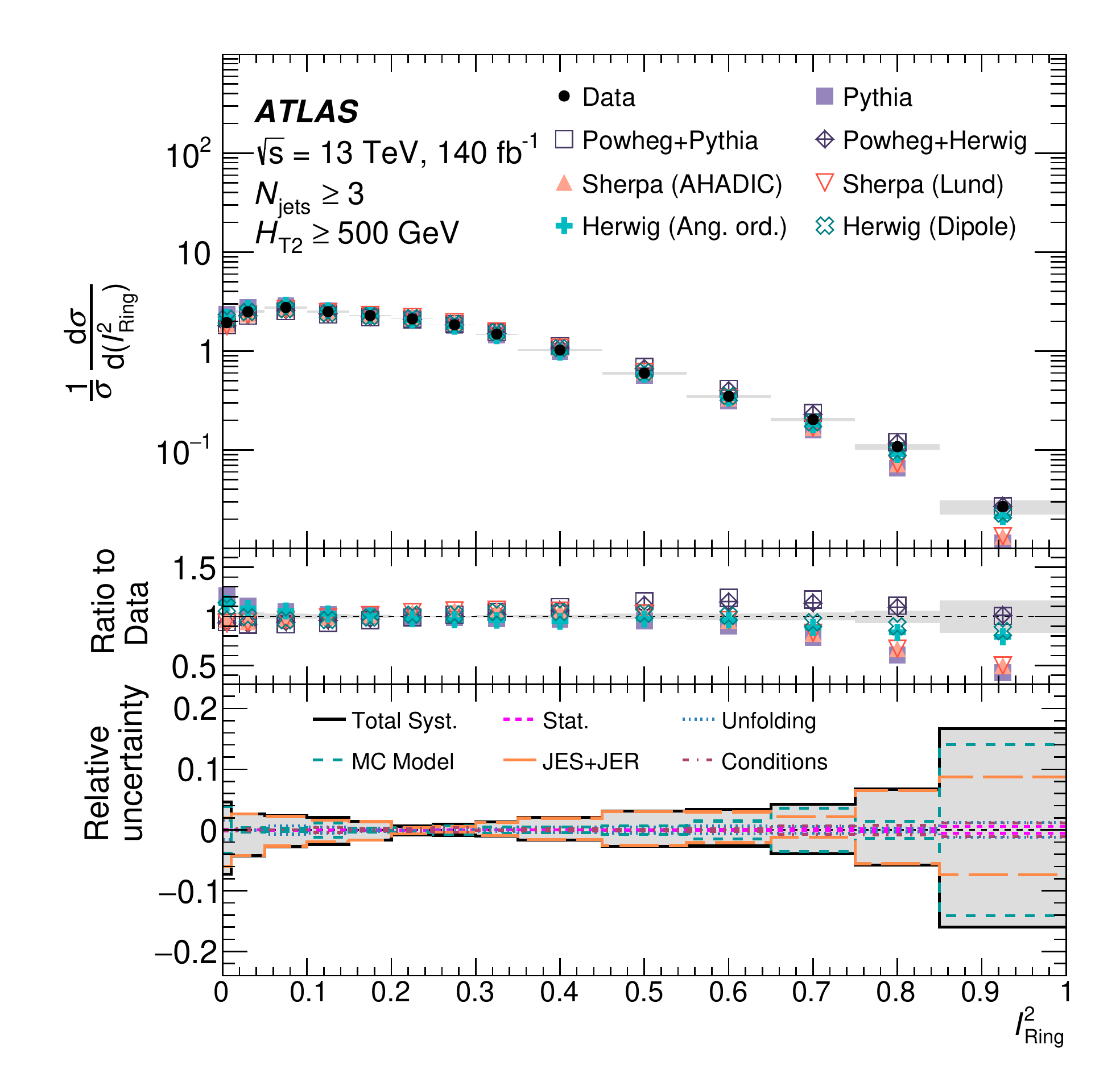}}
\subfigure[$\njets\geq4$]{\includegraphics[width=0.45\textwidth,trim=25 0 25 25,clip]{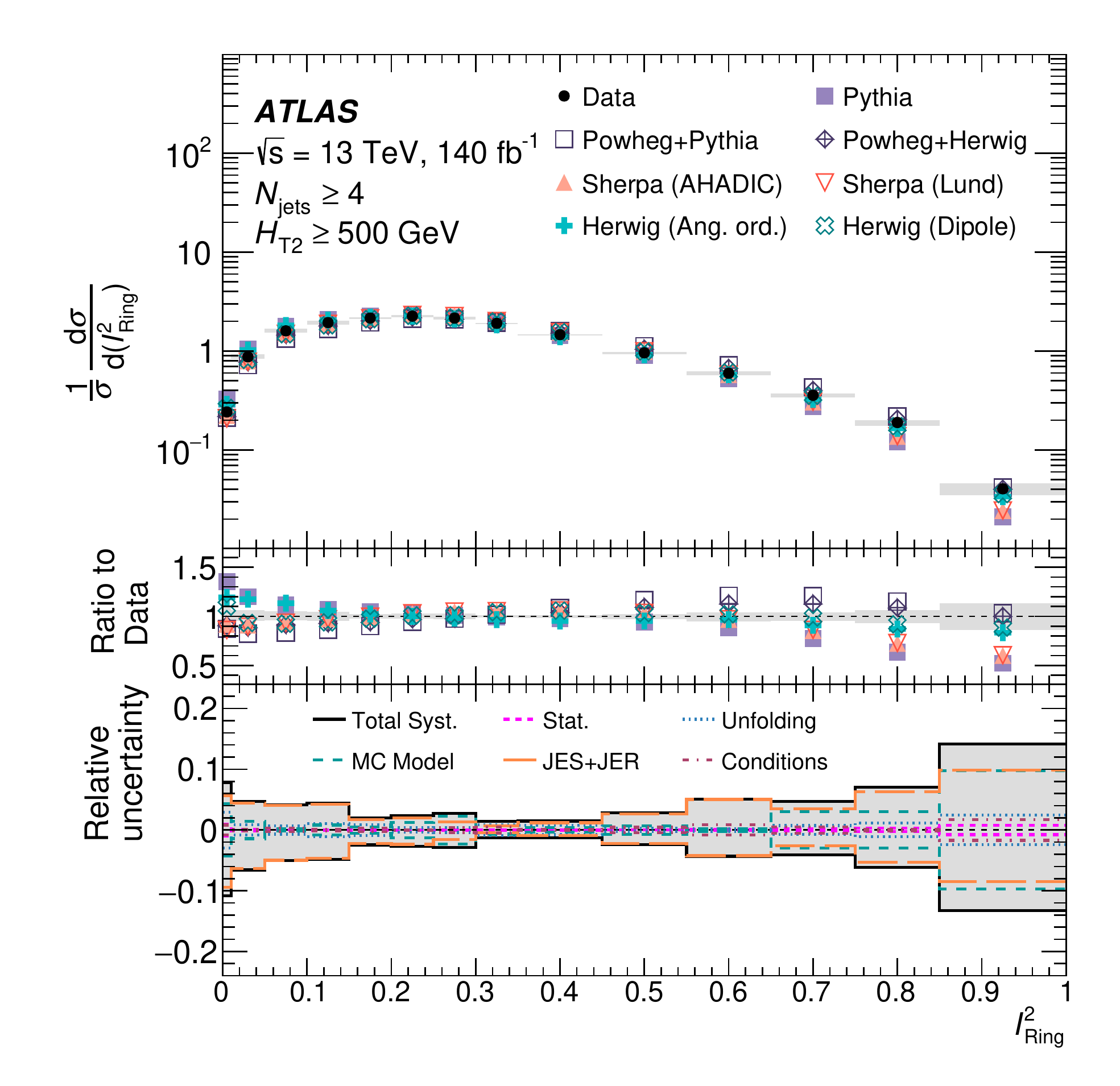}}
\subfigure[$\njets\geq5$]{\includegraphics[width=0.45\textwidth,trim=25 0 25 25,clip]{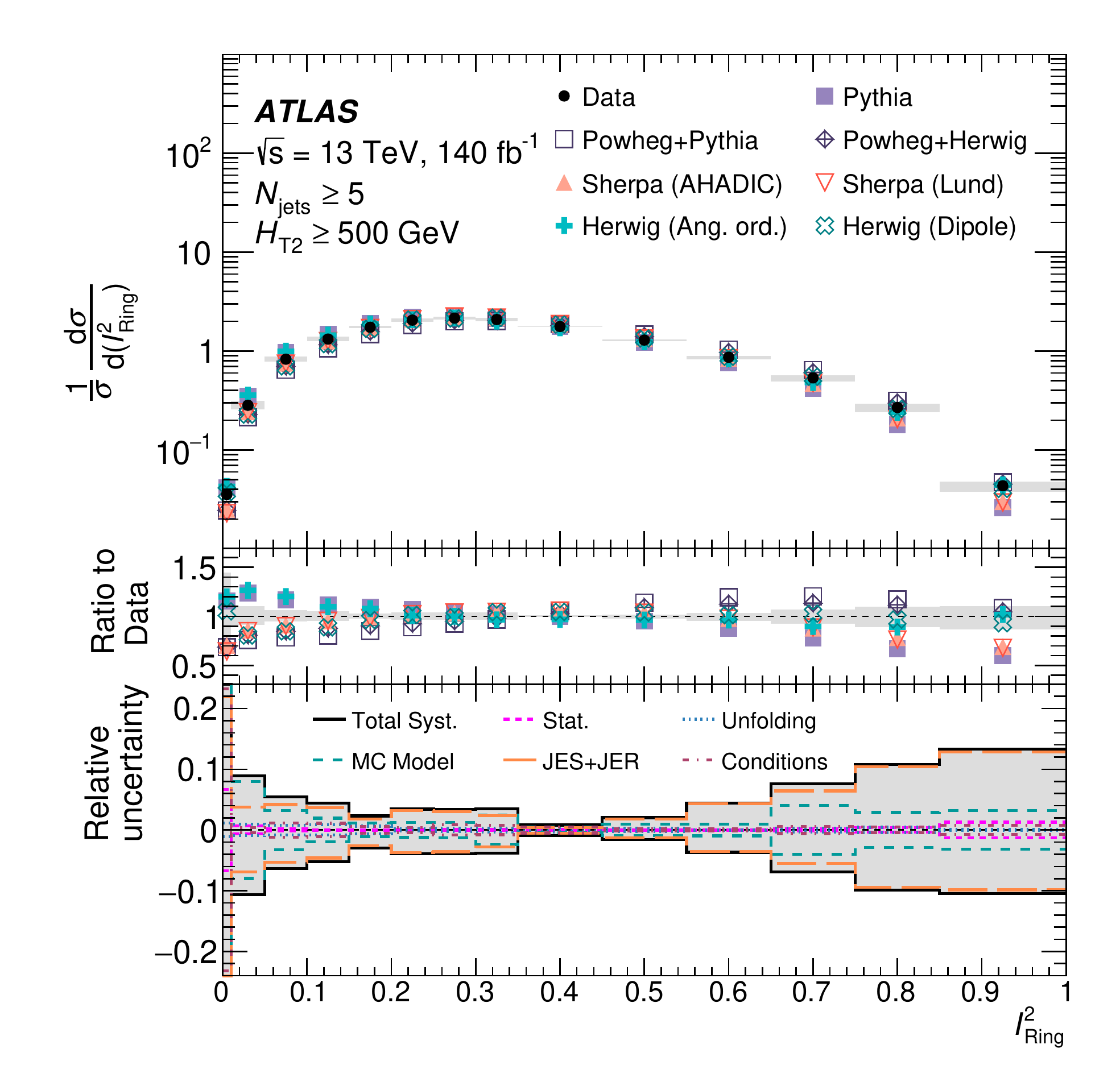}}
\caption{The shape-normalized~\IsoRingTwo~cross-section in data (closed circles), compared with predictions from several Monte Carlo generators.
The distribution is presented for events with $\HTTwo\geq500~\GeV$, in several inclusive bins of \njets.
The middle panels display the ratios of different event generator predictions to the unfolded data.
Event generator predictions are displayed as different marker styles.
The grey band in the upper and middle panels indicates the total uncertainty of the measurement.
The lower panels summarise the various sources of systematic uncertainty in the measurement.
Systematic uncertainties are summarized in groups, with different line styles.
The total uncertainty is shown as a solid black line.
}\label{fig:results:ison:iring2}
\end{figure}
 
\begin{figure}[htbp]
\centering
 
\subfigure[$\njets\geq2$]{\includegraphics[width=0.45\textwidth,trim=25 0 25 25,clip]{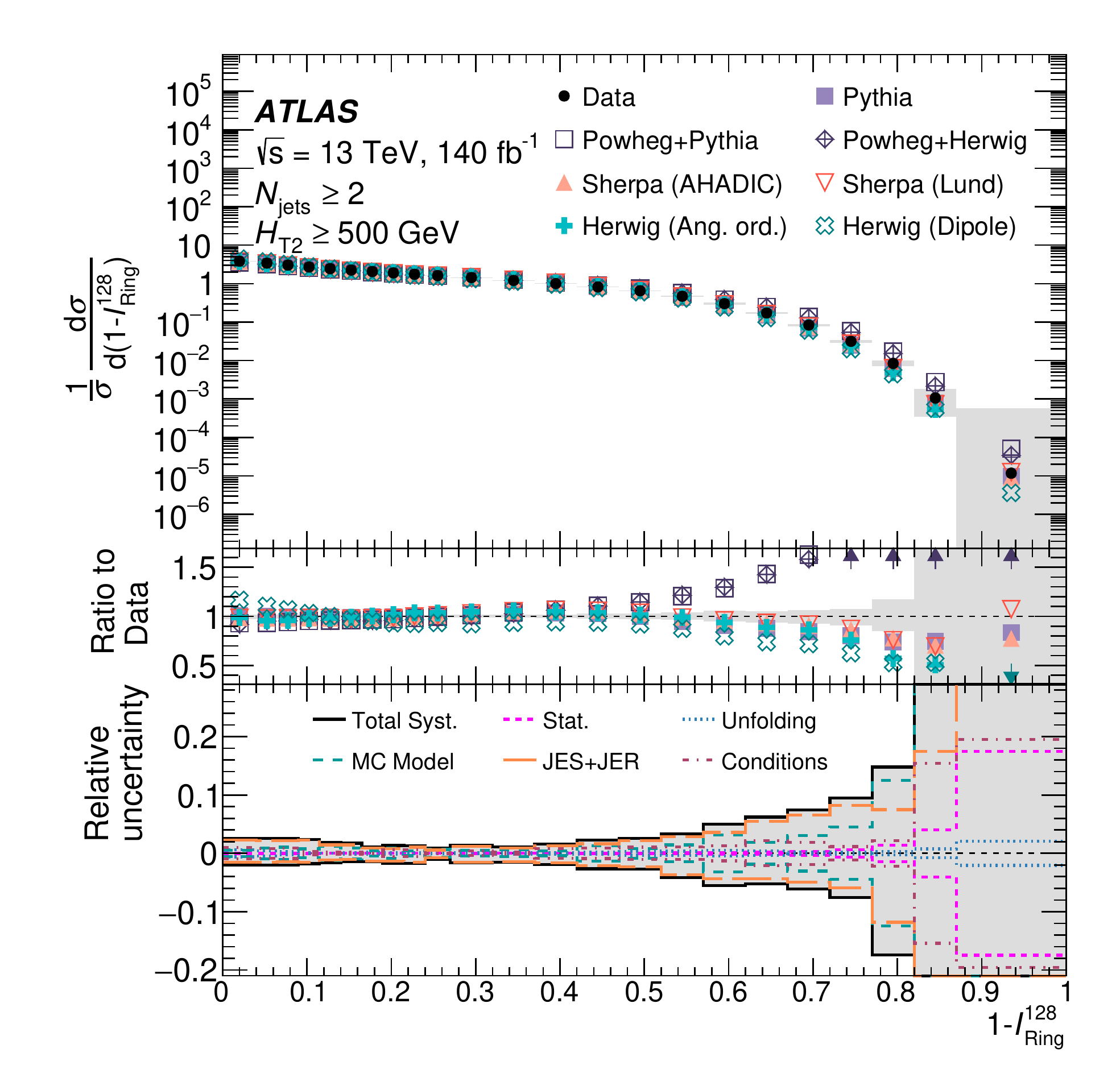}}
\subfigure[$\njets\geq3$]{\includegraphics[width=0.45\textwidth,trim=25 0 25 25,clip]{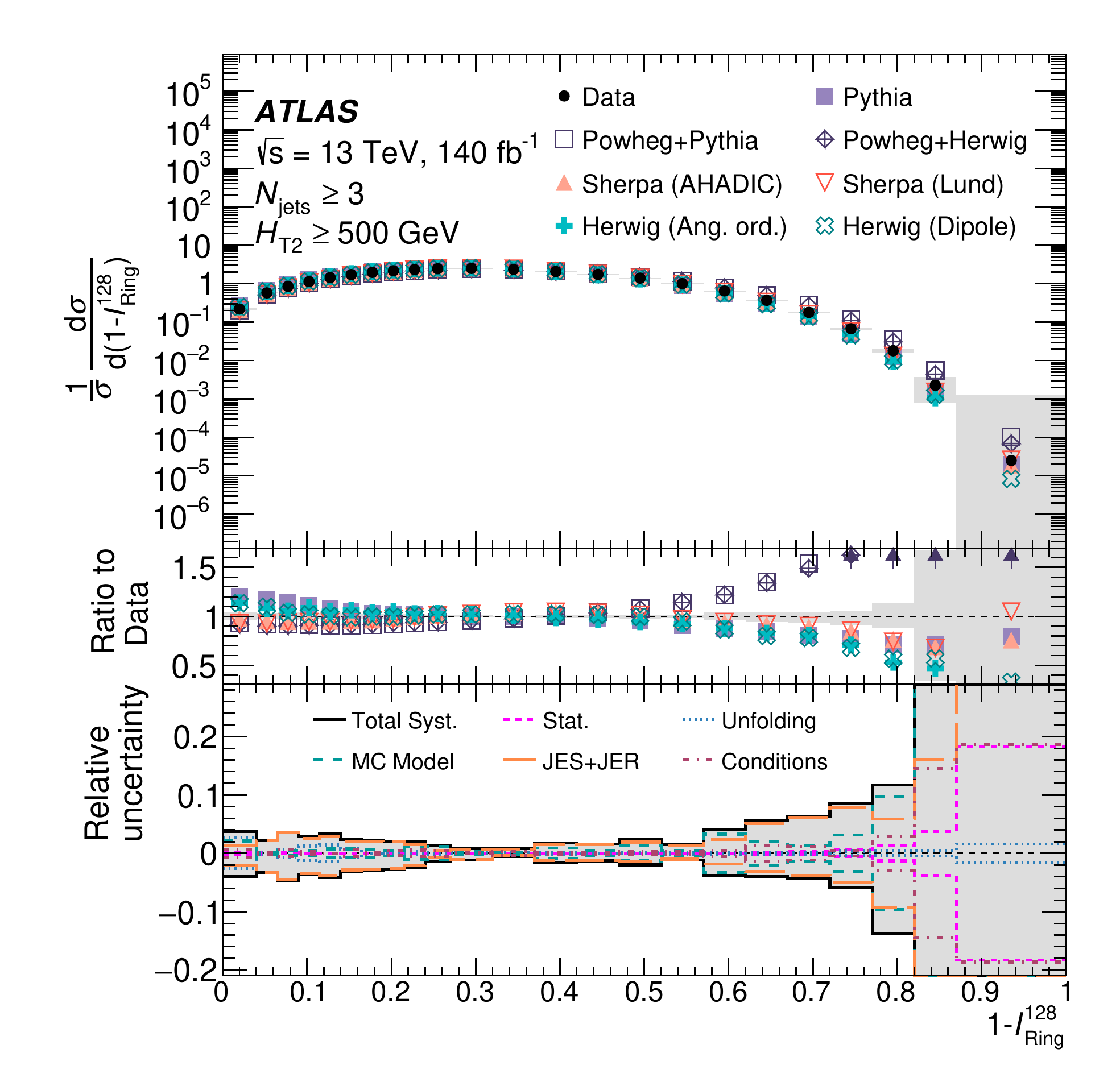}}
\subfigure[$\njets\geq4$]{\includegraphics[width=0.45\textwidth,trim=25 0 25 25,clip]{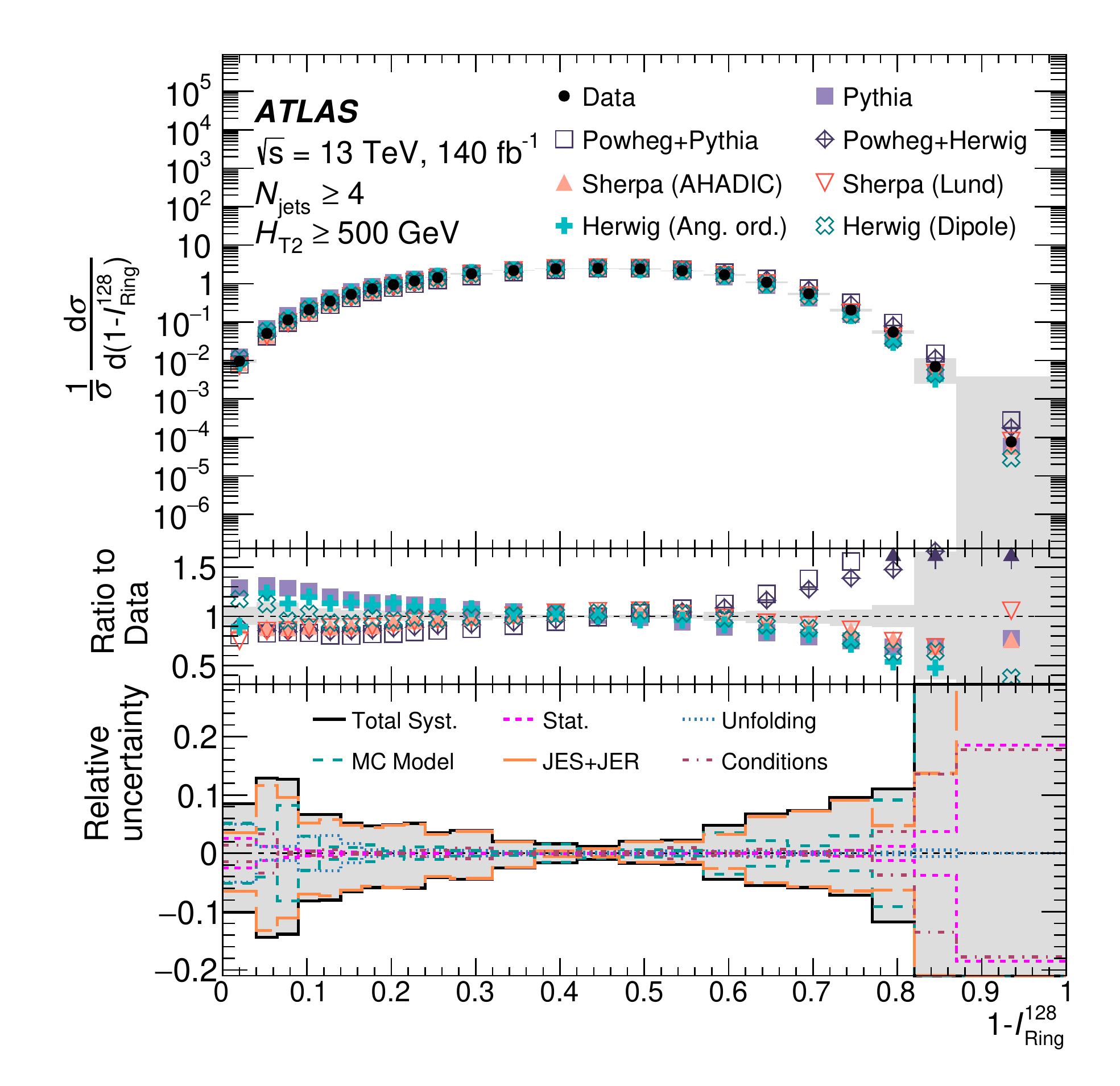}}
\subfigure[$\njets\geq5$]{\includegraphics[width=0.45\textwidth,trim=25 0 25 25,clip]{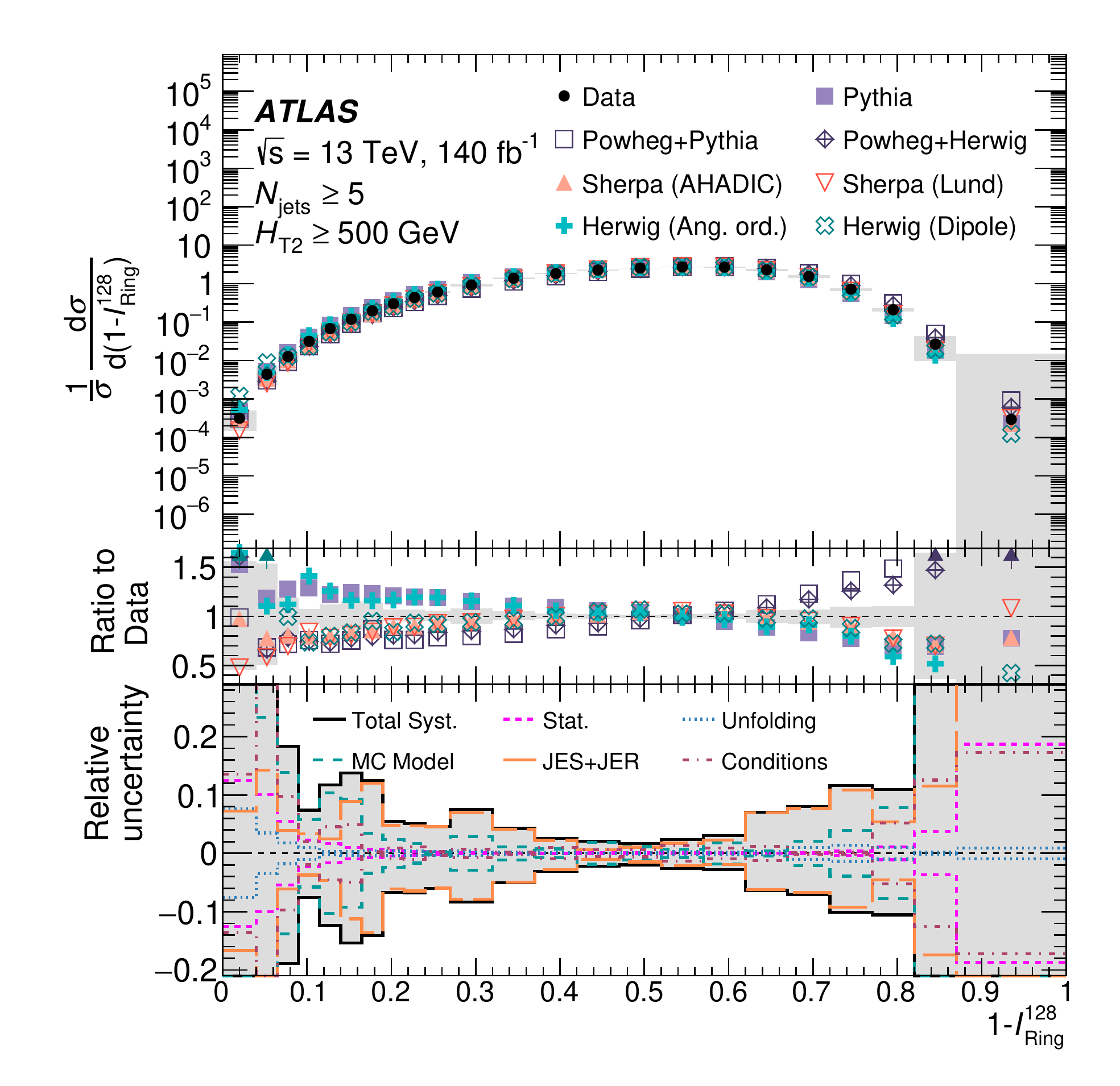}}
 
\caption{The shape-normalized~\IsoRingOneTwentyEight~cross-section in data (closed circles), compared with predictions from several Monte Carlo generators.
The distribution is presented for events with $\HTTwo\geq500~\GeV$, in several inclusive bins of \njets.
The middle panels display the ratios of different event generator predictions to the unfolded data.
The grey band in the upper and middle panels indicates the total uncertainty of the measurement.
If the ratio of a prediction to the unfolded data is outside the range of values displayed in the middle panels, an arrow is drawn at the edge of the panel as an indicator.
Event generator predictions are displayed as different marker styles.
The lower panels summarise the various sources of systematic uncertainty in the measurement.
Systematic uncertainties are summarized in groups, with different line styles.
The total uncertainty is shown as a solid black line.
Some uncertainty bands take values outside the range displayed in the figure: the range is selected for maximum clarity in the bulk of the distribution.
}
\end{figure}
 
\begin{figure}[htbp]
\centering
 
\subfigure[$\njets\geq2$]{\includegraphics[width=0.45\textwidth,trim=25 0 25 25,clip]{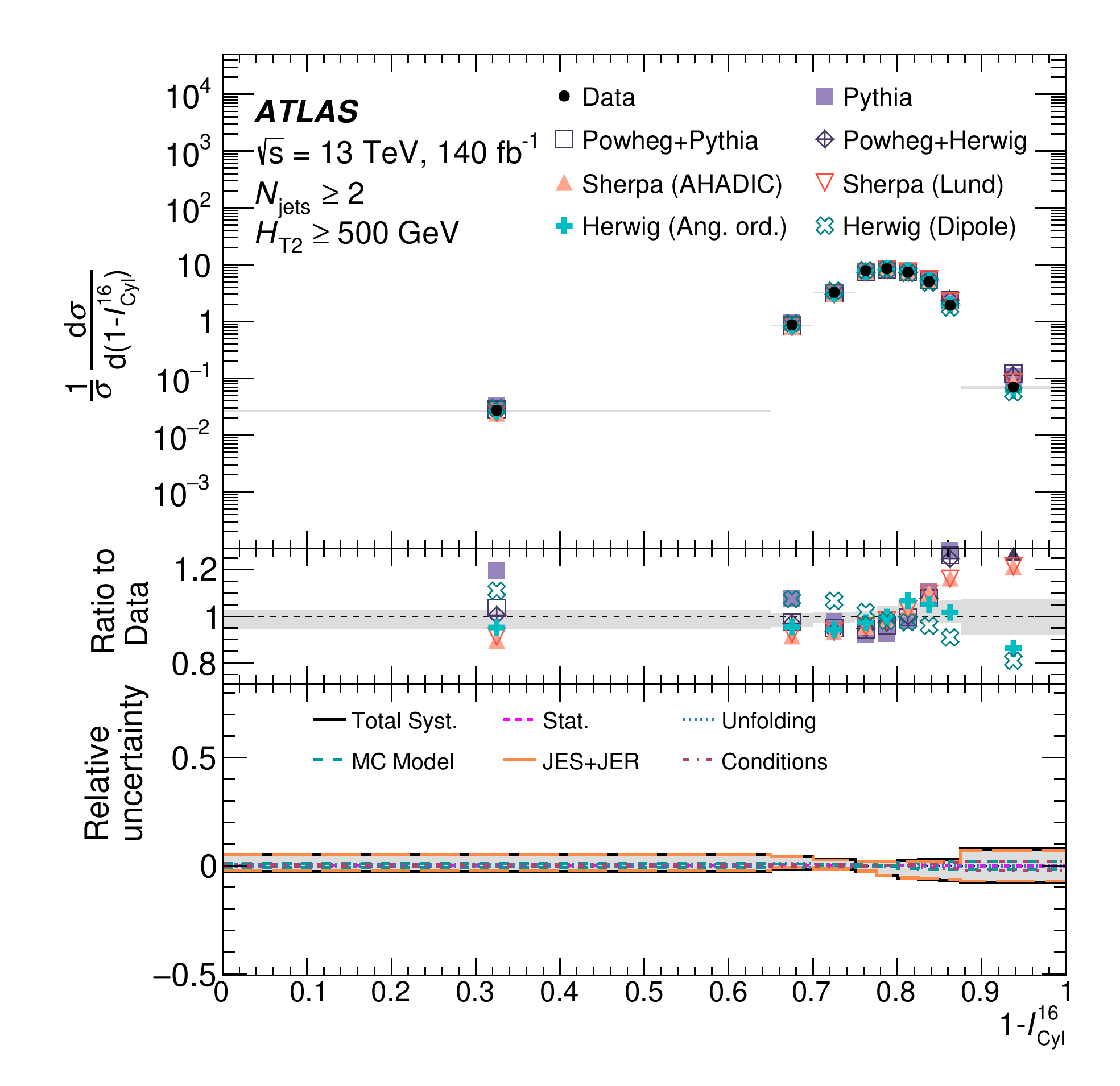}}
\subfigure[$\njets\geq3$]{\includegraphics[width=0.45\textwidth,trim=25 0 25 25,clip]{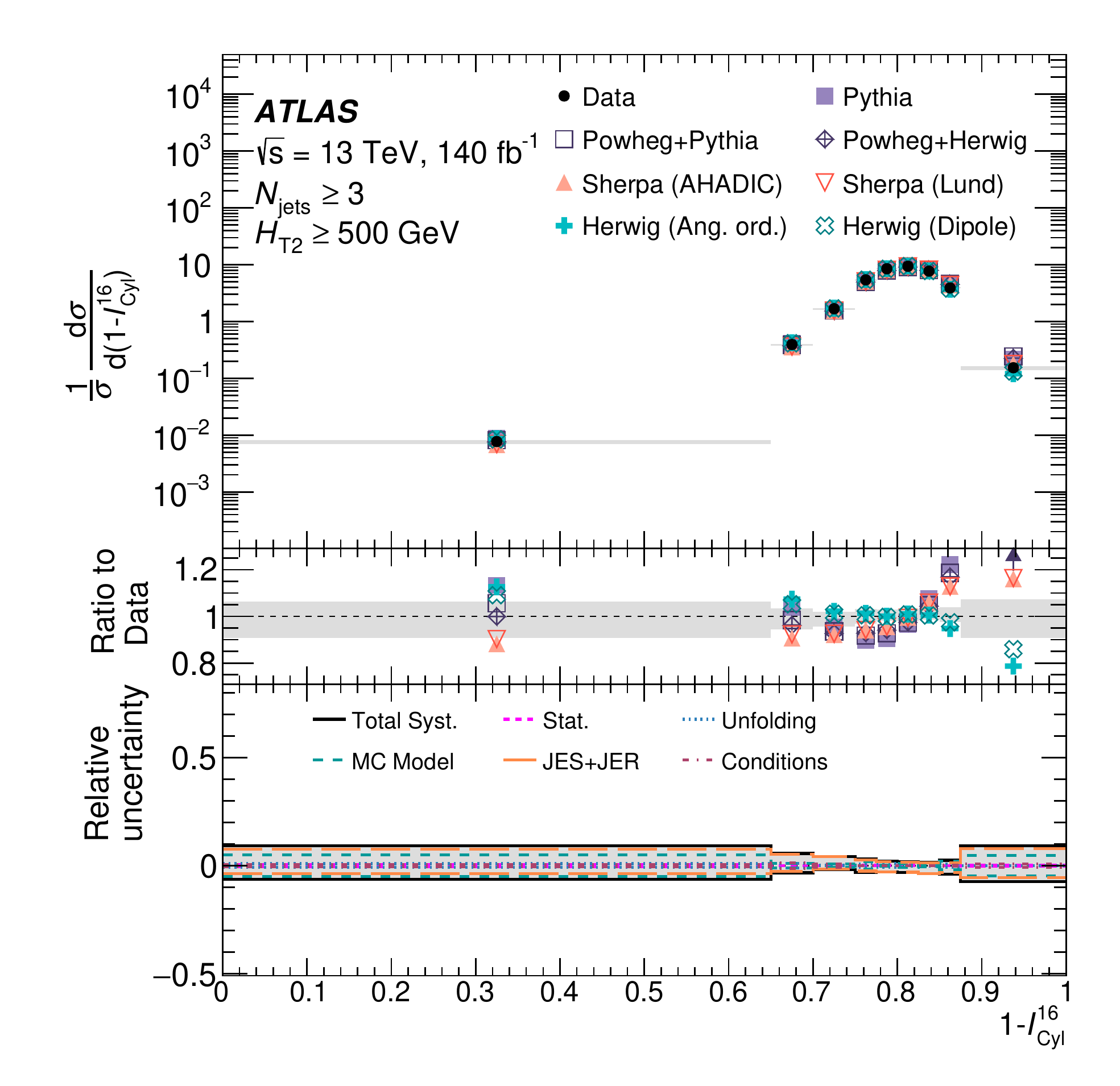}}
\subfigure[$\njets\geq4$]{\includegraphics[width=0.45\textwidth,trim=25 0 25 25,clip]{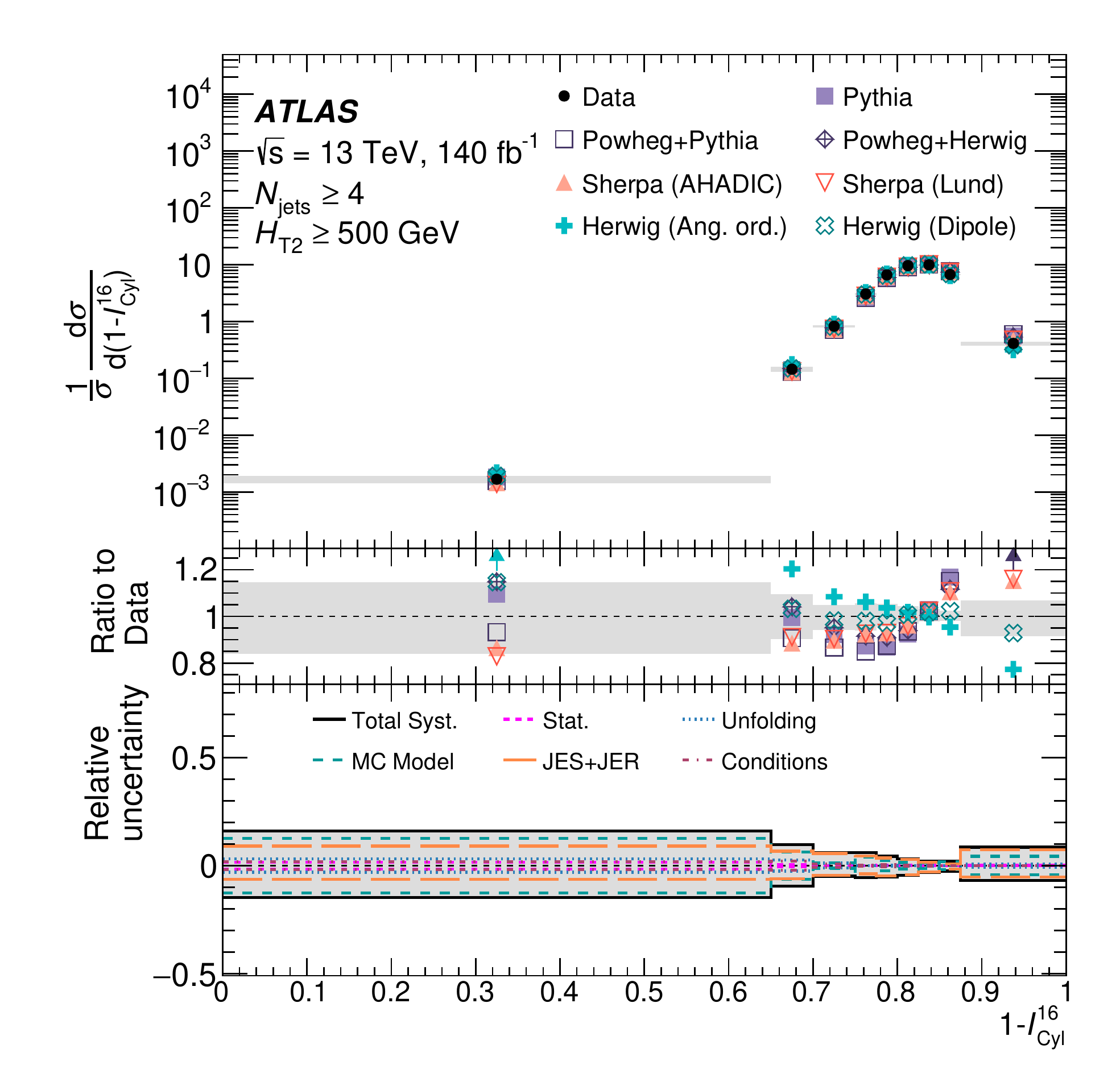}}
\subfigure[$\njets\geq5$]{\includegraphics[width=0.45\textwidth,trim=25 0 25 25,clip]{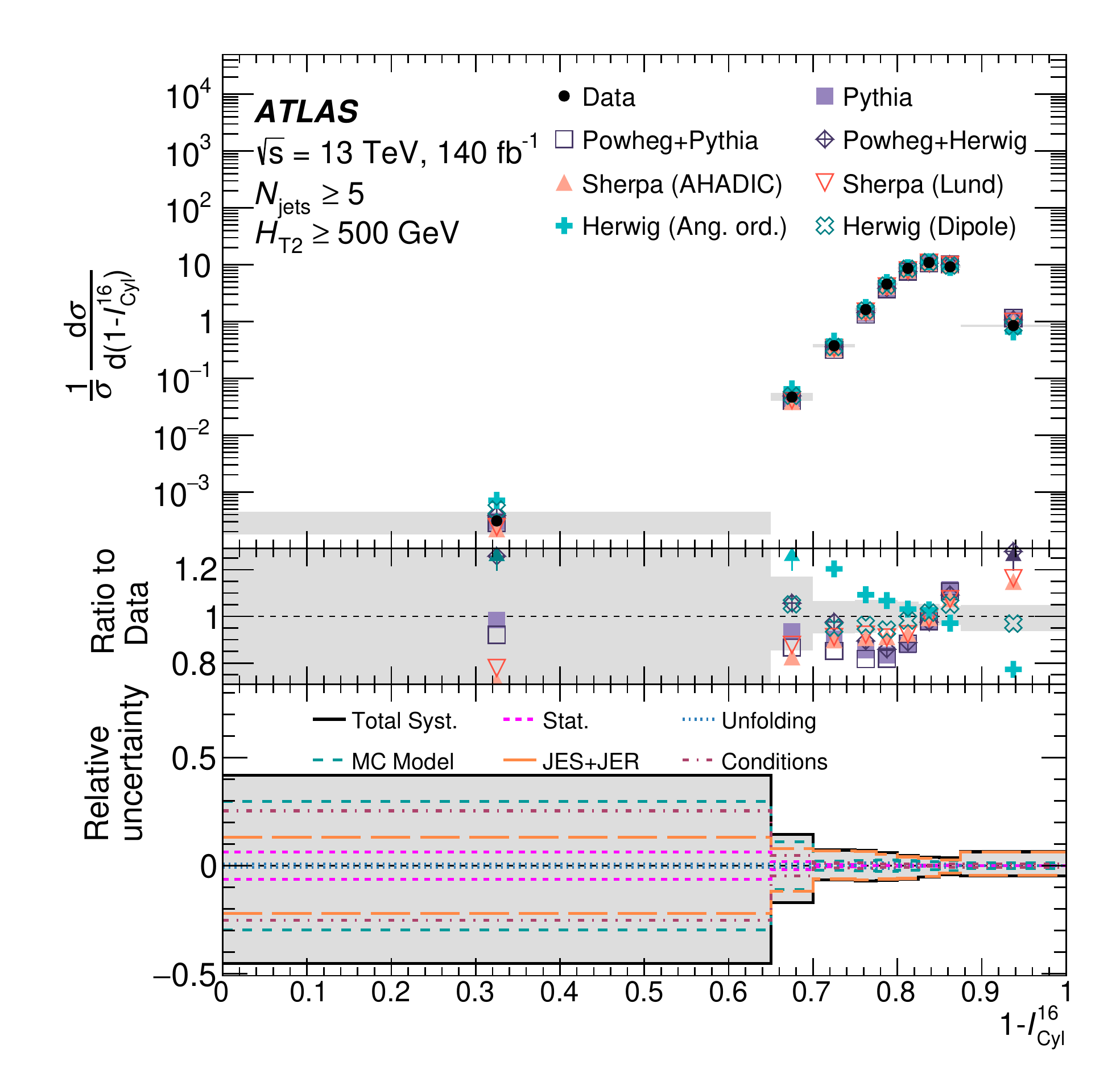}}
 
\caption{The shape-normalized~\IsoCylSixteen~cross-section in data (closed circles), compared with predictions from several Monte Carlo generators.
The distribution is presented for events with $\HTTwo\geq500~\GeV$, in several inclusive bins of \njets.
The middle panels display the ratios of different event generator predictions to the unfolded data.
Event generator predictions are displayed as different marker styles.
The grey band in the upper and middle panels indicates the total uncertainty of the measurement.
If the ratio of a prediction to the unfolded data is outside the range of values displayed in the middle panels, an arrow is drawn at the edge of the panel as an indicator.
The lower panels summarise the various sources of systematic uncertainty in the measurement.
Systematic uncertainties are summarized in groups, with different line styles.
The total uncertainty is shown as a solid black line.
}\label{fig:results:ison:icyl16}
\end{figure}

Finally, cross-sections measured differentially with respect to the event isotropy observables are presented in inclusive bins of both~\njets~and~\HTTwo~for the ring-like isotropies in Figure~\ref{fig:results:isoht2:iring} and for the cylindrical isotropy in Figure~\ref{fig:results:isoht2:icyl16}.
In these figures, events with $\njets\geq5$~are shown in three inclusive $\HTTwo$ bins.
The trends observed are also generally observable for other jet multiplicities.
There are no significant trends in MC modelling that evolve as a function of~\HTTwo.
Events are noted to very gradually become more collimated and dijet-like as the energy scale of the events increases.
The MC predictions' description of the measured data often improves as~\HTTwo~increases, but trends in modelling are similar to those observed in the other measured distributions.
For these triple-differential measurements, the uncertainty that dominates depends on the bin.
At low energy scales, it is typically related to the choice of nominal MC model used in the unfolding procedure.
In the higher $\HTTwo$ bins, the impact of the JES/JER uncertainty on the steeply falling $\HTTwo$ spectrum compounds as the energy scale is increased beyond the region where the measurement is normalized, resulting in degraded precision at high energies.
 
\begin{figure}[htbp]
\centering
 
\subfigure[ \IsoRingTwo, $\HTTwo \geq 500~\GeV$, $\njets\geq5$]{\includegraphics[width=0.32\textwidth,trim=25 0 25 25,clip]{fig_12d.pdf}}
\subfigure[ \IsoRingTwo, $\HTTwo \geq 1000~\GeV$, $\njets\geq5$]{\includegraphics[width=0.32\textwidth,trim=25 0 25 25,clip]{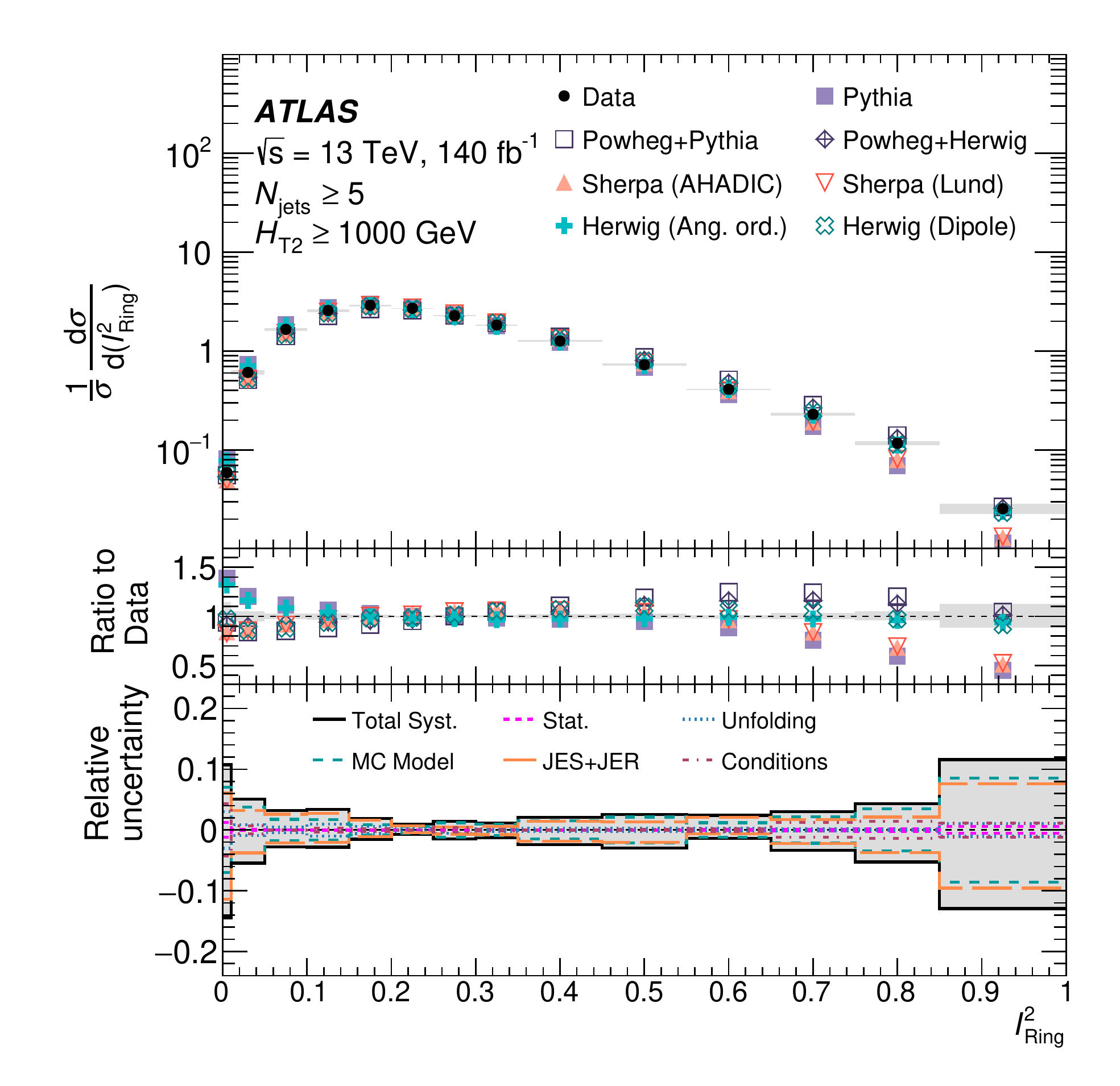}}
\subfigure[ \IsoRingTwo, $\HTTwo \geq 1500~\GeV$, $\njets\geq5$]{\includegraphics[width=0.32\textwidth,trim=25 0 25 25,clip]{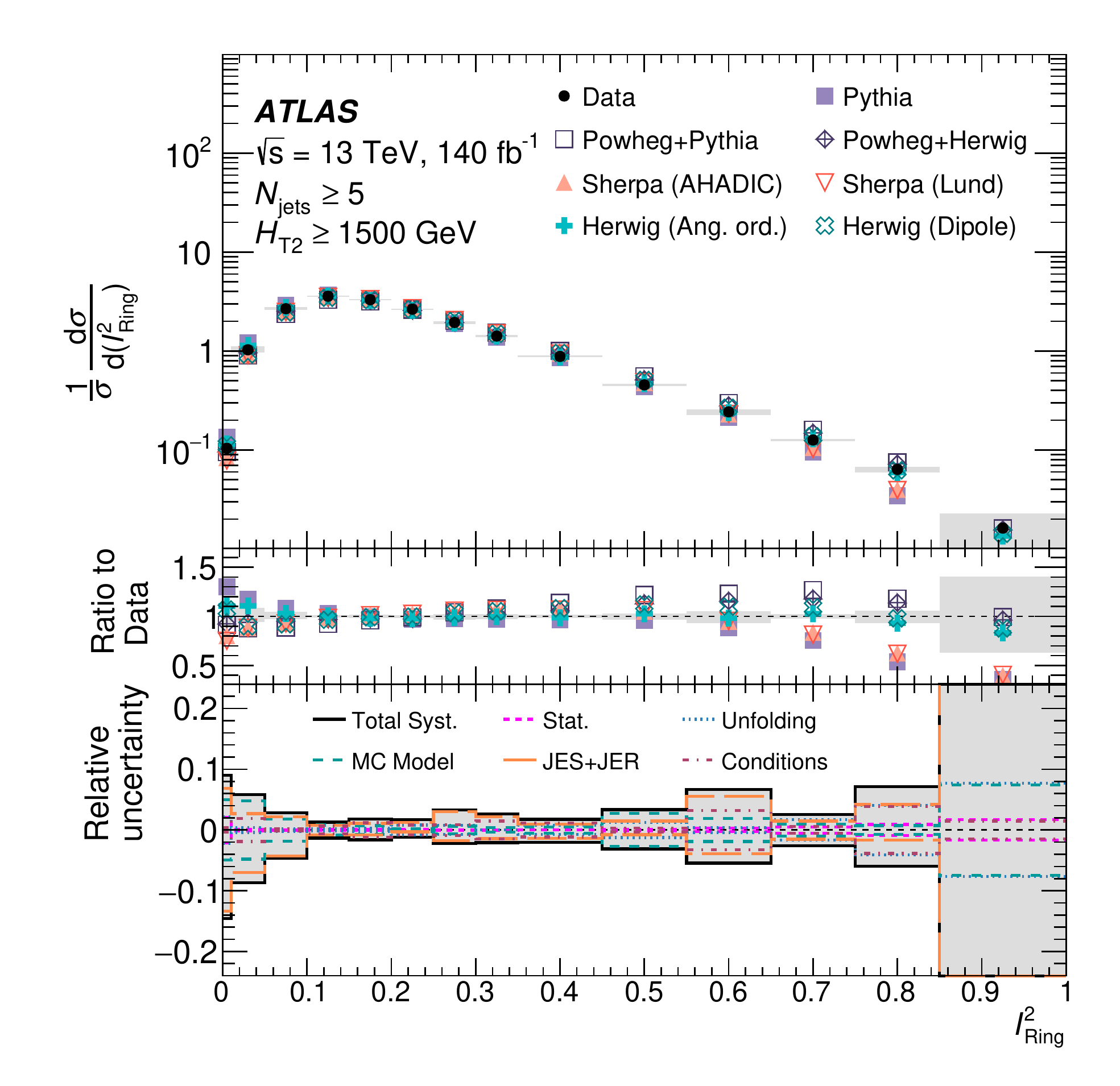}}
 
\subfigure[ \IsoRingOneTwentyEight, $\HTTwo \geq 500~\GeV$, $\njets\geq5$]{\includegraphics[width=0.32\textwidth,trim=25 0 25 25,clip]{fig_13d.pdf}}
\subfigure[ \IsoRingOneTwentyEight, $\HTTwo \geq 1000~\GeV$, $\njets\geq5$]{\includegraphics[width=0.32\textwidth,trim=25 0 25 25,clip]{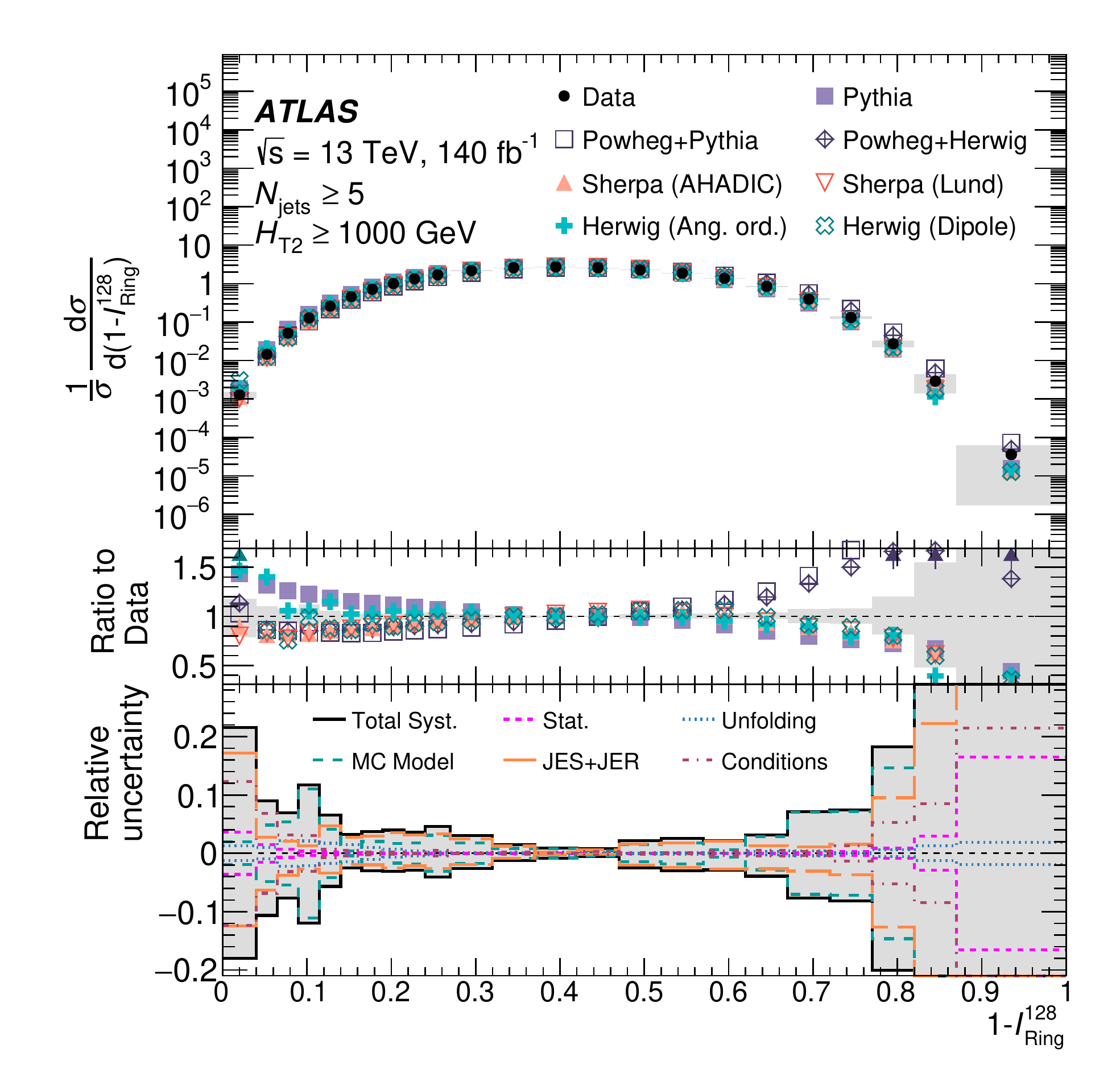}}
\subfigure[ \IsoRingOneTwentyEight, $\HTTwo \geq 1500~\GeV$, $\njets\geq5$]{\includegraphics[width=0.32\textwidth,trim=25 0 25 25,clip]{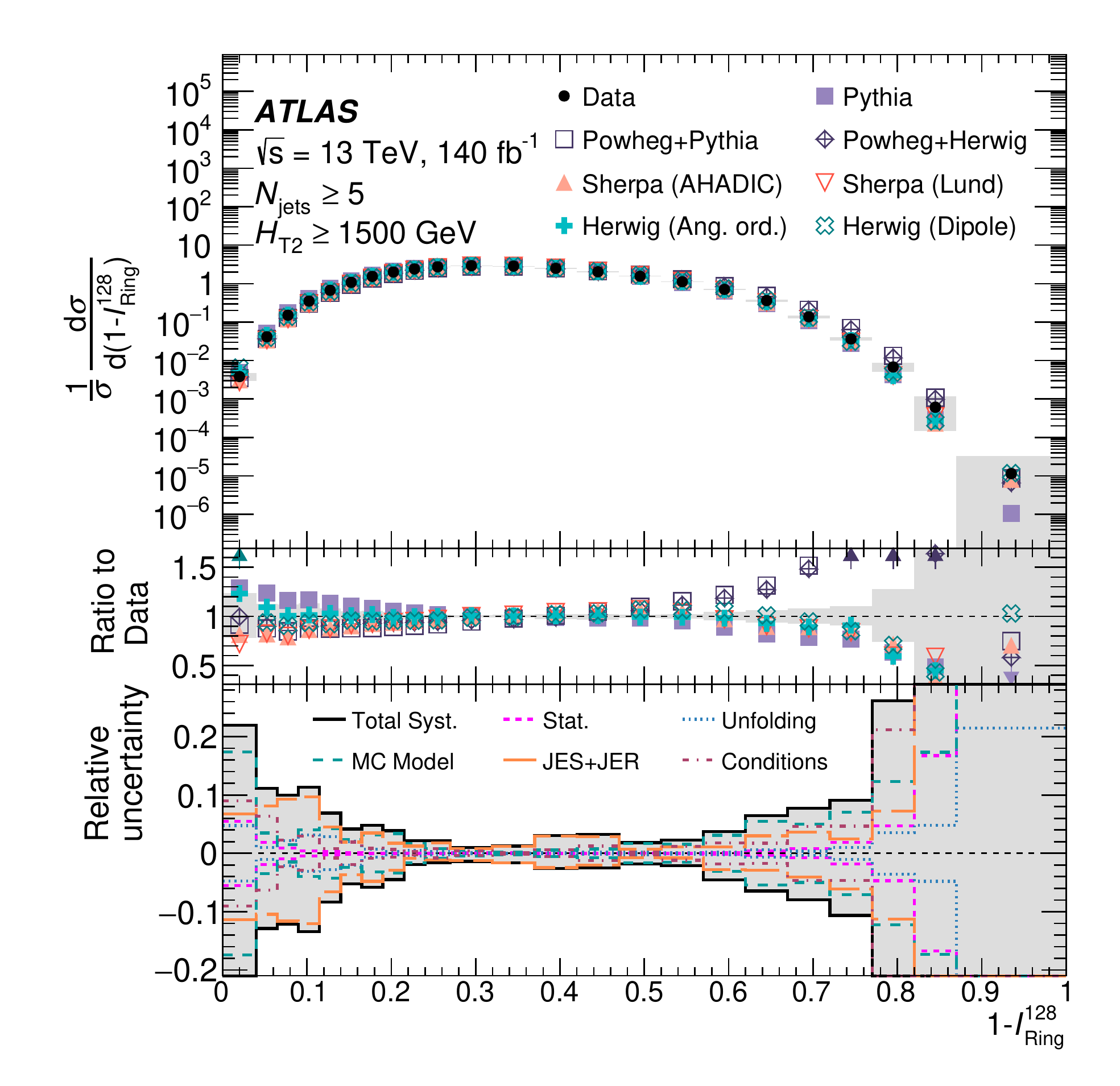}}
 
\caption{The shape-normalized~\IsoRingTwo~and~\IsoRingOneTwentyEight~cross-sections in data (closed circles), compared with predictions from several Monte Carlo generators.
Events with $\njets\geq5$ are presented differentially in inclusive bins of \HTTwo.
The middle panels display the ratios of different event generator predictions to the unfolded data.
Event generator predictions are displayed as different marker styles.
The grey band in the upper and middle panels indicates the total uncertainty of the measurement.
If the ratio of a prediction to the unfolded data is outside the range of values displayed in the middle panels, an arrow is drawn at the edge of the panel as an indicator.
The lower panels summarise the various sources of systematic uncertainty in the measurement.
Systematic uncertainties are summarized in groups, with different line styles.
The total uncertainty is shown as a solid black line.
Some uncertainty bands take values outside the range displayed in the figure: the range is selected for maximum clarity in the bulk of the distribution.
}\label{fig:results:isoht2:iring}
\end{figure}

\begin{figure}[htbp]
\centering
 
\subfigure[ $\HTTwo \geq 500~\GeV$, $\njets\geq5$]{\includegraphics[width=0.32\textwidth,trim=25 0 25 25,clip]{fig_14d.pdf}}
\subfigure[ $\HTTwo \geq 1000~\GeV$, $\njets\geq5$]{\includegraphics[width=0.32\textwidth,trim=25 0 25 25,clip]{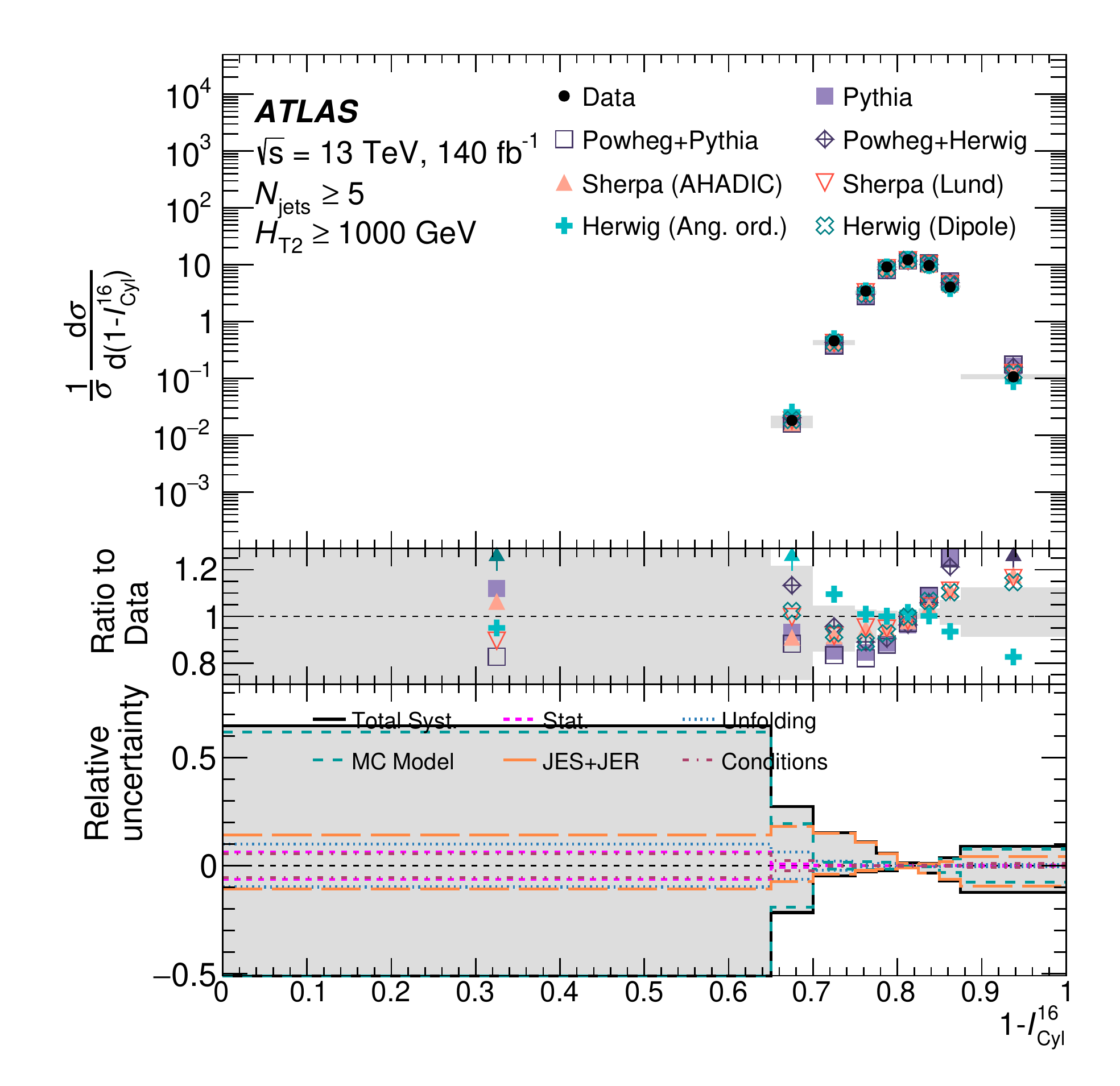}}
\subfigure[ $\HTTwo \geq 1500~\GeV$, $\njets\geq5$]{\includegraphics[width=0.32\textwidth,trim=25 0 25 25,clip]{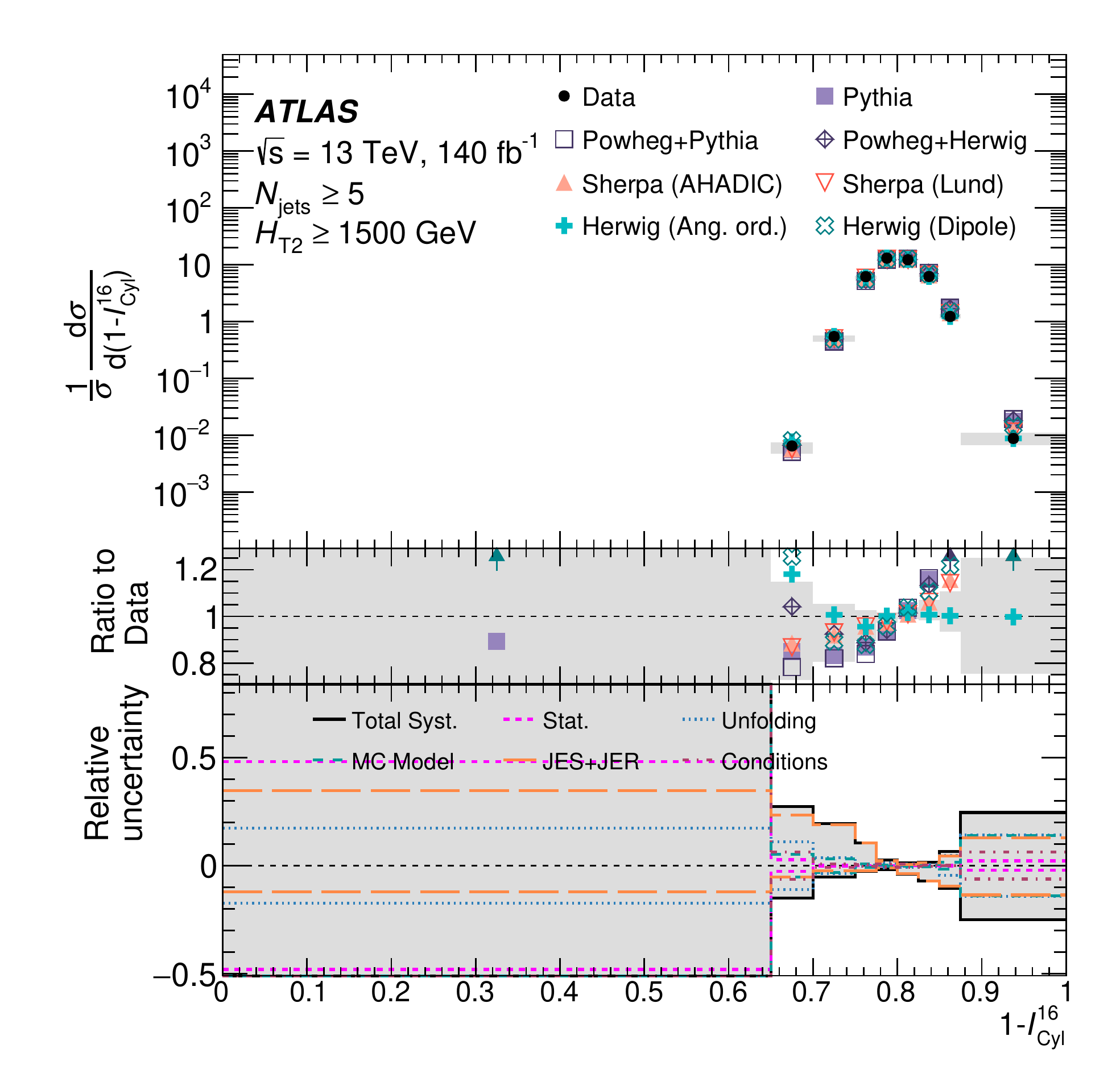}}
 
\caption{The shape-normalised~\IsoCylSixteen~cross-section in data (closed circles), compared with predictions from several Monte Carlo generators.
Events with $\njets\geq5$ are presented differentially in inclusive bins of \HTTwo.
The middle panels display the ratios of different event generator predictions to the unfolded data.
Event generator predictions are displayed as different marker styles.
The grey band in the upper and middle panels indicates the total uncertainty of the measurement.
If the ratio of a prediction to the unfolded data is outside the range of values displayed in the middle panels, an arrow is drawn at the edge of the panel as an indicator.
The lower panels summarise the various sources of systematic uncertainty in the measurement.
Systematic uncertainties are summarized in groups, with different line styles.
The total uncertainty is shown as a solid black line.
Some uncertainty bands take values outside the range displayed in the figure: the range is selected for maximum clarity in the bulk of the distribution.
}\label{fig:results:isoht2:icyl16}
\end{figure}


\FloatBarrier
 
\section{Concluding remarks}

A measurement of novel event-shape observables that describe collider events in terms of their \emph{event isotropy} has been performed in 139~\ifb\ of proton--proton collisions with centre-of-mass energy $\sqrt{s}=13~\TeV$, recorded with the ATLAS detector at CERN's Large Hadron Collider.
These event shapes are defined in terms of isotropic reference geometries with cylindrical and circular symmetries, using the Energy-Mover's Distance to quantify Wasserstein distances between multijet events and the isotropic configurations in terms of optimal transport problems.
Event isotropies are shown to have increased sensitivity to isotropic multijet events when compared with other event shapes such as the transverse thrust.
They are capable of exposing a remote piece of QCD phase space that is difficult to model and relevant to many searches for physics beyond the Standard Model.
 
Cross-sections are measured differentially with respect to three event-isotropy observables in inclusive bins of jet multiplicity and \HTTwo.
These measurements are corrected for acceptance and detector resolution effects, and normalised relative to the number of events passing the analysis selection in each such bin.
This procedure allows the measurement to directly probe the shape of the event isotropies.
 
The measured data are compared with the predictions of several state-of-the-art Monte Carlo event generators.
Agreement between the unfolded data and the simulated events tends to be best in balanced, dijet-like arrangements and deteriorates in more isotropic configurations.
For the measurement of~\IsoRingTwo, an observable that interpolates between balanced dijet and trijet events similarly to the transverse thrust, the predictions of NLO MC generators generally outperform those of LO simulation.
In the measurement of $1-\IsoRingOneTwentyEight$, which interpolates between balanced dijet events and isotropic multijet configurations in the transverse plane, no single event generator accurately describes the distribution.
In particular, the descriptions from the NLO~\POWPY and \HERWIG~simulations differ in the region sensitive to isotropic configurations.
The two-dimensional isotropy $1-\IsoCylSixteen$ interpolates between forward dijet events and multijet events with activity more evenly covering the rapidity--azimuth plane.
This observable is not well-predicted by any MC generator, and elicits large differences between the parton shower models available in~\HERWIG.
 
A Rivet routine is available for this measurement~\cite{Buckley:2010ar}, and the measured data points have been made publicly available along with other auxiliary information~\cite{hepdata.110164} for use in future Monte Carlo tuning campaigns and other studies of QCD.


\section*{Acknowledgements}


We thank CERN for the very successful operation of the LHC, as well as the
support staff from our institutions without whom ATLAS could not be
operated efficiently.
 
We acknowledge the support of
ANPCyT, Argentina;
YerPhI, Armenia;
ARC, Australia;
BMWFW and FWF, Austria;
ANAS, Azerbaijan;
CNPq and FAPESP, Brazil;
NSERC, NRC and CFI, Canada;
CERN;
ANID, Chile;
CAS, MOST and NSFC, China;
Minciencias, Colombia;
MEYS CR, Czech Republic;
DNRF and DNSRC, Denmark;
IN2P3-CNRS and CEA-DRF/IRFU, France;
SRNSFG, Georgia;
BMBF, HGF and MPG, Germany;
GSRI, Greece;
RGC and Hong Kong SAR, China;
ISF and Benoziyo Center, Israel;
INFN, Italy;
MEXT and JSPS, Japan;
CNRST, Morocco;
NWO, Netherlands;
RCN, Norway;
MEiN, Poland;
FCT, Portugal;
MNE/IFA, Romania;
MESTD, Serbia;
MSSR, Slovakia;
ARRS and MIZ\v{S}, Slovenia;
DSI/NRF, South Africa;
MICINN, Spain;
SRC and Wallenberg Foundation, Sweden;
SERI, SNSF and Cantons of Bern and Geneva, Switzerland;
MOST, Taiwan;
TENMAK, T\"urkiye;
STFC, United Kingdom;
DOE and NSF, United States of America.
In addition, individual groups and members have received support from
BCKDF, CANARIE, Compute Canada and CRC, Canada;
PRIMUS 21/SCI/017 and UNCE SCI/013, Czech Republic;
COST, ERC, ERDF, Horizon 2020 and Marie Sk{\l}odowska-Curie Actions, European Union;
Investissements d'Avenir Labex, Investissements d'Avenir Idex and ANR, France;
DFG and AvH Foundation, Germany;
Herakleitos, Thales and Aristeia programmes co-financed by EU-ESF and the Greek NSRF, Greece;
BSF-NSF and MINERVA, Israel;
Norwegian Financial Mechanism 2014-2021, Norway;
NCN and NAWA, Poland;
La Caixa Banking Foundation, CERCA Programme Generalitat de Catalunya and PROMETEO and GenT Programmes Generalitat Valenciana, Spain;
G\"{o}ran Gustafssons Stiftelse, Sweden;
The Royal Society and Leverhulme Trust, United Kingdom.
 
The crucial computing support from all WLCG partners is acknowledged gratefully, in particular from CERN, the ATLAS Tier-1 facilities at TRIUMF (Canada), NDGF (Denmark, Norway, Sweden), CC-IN2P3 (France), KIT/GridKA (Germany), INFN-CNAF (Italy), NL-T1 (Netherlands), PIC (Spain), ASGC (Taiwan), RAL (UK) and BNL (USA), the Tier-2 facilities worldwide and large non-WLCG resource providers. Major contributors of computing resources are listed in Ref.~\cite{ATL-SOFT-PUB-2021-003}.


\printbibliography
 
\clearpage
 
\begin{flushleft}
\hypersetup{urlcolor=black}
{\Large The ATLAS Collaboration}

\bigskip

\AtlasOrcid[0000-0002-6665-4934]{G.~Aad}$^\textrm{\scriptsize 102}$,
\AtlasOrcid[0000-0002-5888-2734]{B.~Abbott}$^\textrm{\scriptsize 120}$,
\AtlasOrcid[0000-0002-7248-3203]{D.C.~Abbott}$^\textrm{\scriptsize 103}$,
\AtlasOrcid[0000-0002-1002-1652]{K.~Abeling}$^\textrm{\scriptsize 55}$,
\AtlasOrcid[0000-0002-8496-9294]{S.H.~Abidi}$^\textrm{\scriptsize 29}$,
\AtlasOrcid[0000-0002-9987-2292]{A.~Aboulhorma}$^\textrm{\scriptsize 35e}$,
\AtlasOrcid[0000-0001-5329-6640]{H.~Abramowicz}$^\textrm{\scriptsize 151}$,
\AtlasOrcid[0000-0002-1599-2896]{H.~Abreu}$^\textrm{\scriptsize 150}$,
\AtlasOrcid[0000-0003-0403-3697]{Y.~Abulaiti}$^\textrm{\scriptsize 117}$,
\AtlasOrcid[0000-0003-0762-7204]{A.C.~Abusleme~Hoffman}$^\textrm{\scriptsize 137a}$
% ...
\bigskip
\\

% ...
$^{29}$Physics Department, Brookhaven National Laboratory, Upton NY; United States of America.\\
% ...
$^{35}$$^{(a)}$Facult\'e des Sciences Ain Chock, R\'eseau Universitaire de Physique des Hautes Energies - Universit\'e Hassan II, Casablanca;$^{(b)}$Facult\'{e} des Sciences, Universit\'{e} Ibn-Tofail, K\'{e}nitra;$^{(c)}$Facult\'e des Sciences Semlalia, Universit\'e Cadi Ayyad, LPHEA-Marrakech;$^{(d)}$LPMR, Facult\'e des Sciences, Universit\'e Mohamed Premier, Oujda;$^{(e)}$Facult\'e des sciences, Universit\'e Mohammed V, Rabat;$^{(f)}$Institute of Applied Physics, Mohammed VI Polytechnic University, Ben Guerir; Morocco.\\
% ...
$^{55}$II. Physikalisches Institut, Georg-August-Universit\"{a}t G\"ottingen, G\"ottingen; Germany.\\
% ...
$^{102}$CPPM, Aix-Marseille Universit\'e, CNRS/IN2P3, Marseille; France.\\
$^{103}$Department of Physics, University of Massachusetts, Amherst MA; United States of America.\\
% ...
$^{117}$Department of Physics, New York University, New York NY; United States of America.\\
% ...
$^{120}$Homer L. Dodge Department of Physics and Astronomy, University of Oklahoma, Norman OK; United States of America.\\
% ...
$^{137}$$^{(a)}$Departamento de F\'isica, Pontificia Universidad Cat\'olica de Chile, Santiago;$^{(b)}$Millennium Institute for Subatomic physics at high energy frontier (SAPHIR), Santiago;$^{(c)}$Instituto de Investigaci\'on Multidisciplinario en Ciencia y Tecnolog\'ia, y Departamento de F\'isica, Universidad de La Serena;$^{(d)}$Universidad Andres Bello, Department of Physics, Santiago;$^{(e)}$Instituto de Alta Investigaci\'on, Universidad de Tarapac\'a, Arica;$^{(f)}$Departamento de F\'isica, Universidad T\'ecnica Federico Santa Mar\'ia, Valpara\'iso; Chile.\\
% ...
$^{150}$Department of Physics, Technion, Israel Institute of Technology, Haifa; Israel.\\
$^{151}$Raymond and Beverly Sackler School of Physics and Astronomy, Tel Aviv University, Tel Aviv; Israel.\\
% ...
$^{a}$ Also Affiliated with an institute covered by a cooperation agreement with CERN.\\
$^{b}$ Also at An-Najah National University, Nablus; Palestine.\\
$^{c}$ Also at Borough of Manhattan Community College, City University of New York, New York NY; United States of America.\\
$^{d}$ Also at Bruno Kessler Foundation, Trento; Italy.\\
$^{e}$ Also at Center for High Energy Physics, Peking University; China.\\
$^{f}$ Also at Center for Interdisciplinary Research and Innovation (CIRI-AUTH), Thessaloniki ; Greece.\\
% ...

\end{flushleft}


\end{document}